\documentclass{aa}
\usepackage{graphicx}
\usepackage{txfonts}
\usepackage[]{natbib}
\newcommand{\spitzer}{{\it Spitzer}}
\newcommand{\galex}{{\it GALEX}}
\newcommand{\mi}{\,$\mu$m}
\newcommand{\av}{{$A_{\rm V}$}}
\newcommand{\hii}{\ion{H}{ii}}
\newcommand{\hi}{\ion{H}{i}}
\newcommand{\mm}{M\,33}

\newcommand{\Ha}{H$\alpha$}
\newcommand{\msun}{\,M$_\odot$}
\newcommand{\lsun}{\,L$_\odot$}

\defcitealias{2007A&A...476.1161V}{Paper I}
\begin{document}
   \title{Star formation in \mm: multiwavelength signatures across the disk}

   \author{S. Verley
          \inst{1}
          \and
          E. Corbelli
          \inst{1}
         \and
          C. Giovanardi
          \inst{1}
          \and
           L. K. Hunt
          \inst{2}
}


   \institute{Osservatorio Astrofisico di Arcetri, Largo E. Fermi, 5 - 50125
Firenze - Italy\\
              \email{[simon, edvige, giova]@arcetri.astro.it}
         \and
              INAF - Istituto di Radioastronomia-Sezione Firenze, Largo E. Fermi 5, 50125 Firenze,
Italy\\
              \email{hunt@arcetri.astro.it}
             }

   \date{Received; accepted}


  \abstract
   {}
{We use different tracers, such as \Ha,
ultraviolet (UV), and infrared (IR) emissions at various wavelengths,
to study the dust and star-formation (SF) conditions throughout the disk
of \mm.}
{We derive the radial distribution of dust, of the old and young stellar
population using \textit{Spitzer} and \textit{GALEX} data, complemented
by ground-based optical data and available surveys of atomic
and molecular gas. We separate the contribution of discrete sources
to the IR brightness from the diffuse emission.}
{At 8 and 24\mi, discrete sources account for $\gtrsim40$\% of the IR emission
in the innermost 3~kpc, and for $\lesssim20$\% further out.
We find that stochastic emission from very small grains in the diffuse
interstellar medium accounts for only $\sim$10\% of the diffuse 24\mi\ emission, 
and that dusty circumstellar shells
of unresolved, evolved AGB stars (carbon stars) are a viable alternative.
The 8\mi\ profile suggests that PAH emission
declines faster with radius than the dust continuum. 
In annular regions, 0.24~kpc wide, we find a mean extinction value for stellar 
continuum $A_{\rm V} \sim$ 0.25\,mag with a weak dependence on radius, consistent with
the shallow metallicity gradient observed. Dust opacity derived from the 160\mi\
emission
decreases instead  by a factor 10 from the center to edge of the SF disk.}
{Using extinction corrected UV and \Ha\ maps we find the global SF rate in \mm,
over the last 100~Myr, to be $0.45\pm 0.10$ M$_\odot$ yr$^{-1}$.
FIR (far-IR) and TIR (total-IR) luminosities can trace SF even though
a high conversion factor is required to recover the effective rate.
If carbon stars are powering the diffuse 24\mi\ emission in \mm\ this
can trace star formation 1~Gyr ago and provide a more complete view of the 
SF history of the galaxy. Today
the SF rate declines radially with a scale length of $\sim2$~kpc, longer than 
for the old stellar population, suggesting an inside-out growth of the disk.}


   \maketitle
%

\section{Introduction}

Our knowledge of the conditions in the interstellar medium (ISM) that favour
star formation (SF) is mostly based on
our Galaxy and on luminous, gas-rich galaxies with high SF rates (SFRs).
Thus, any proposed SF scenario is biased against different Hubble types,
especially against quiescent galaxies with a low molecular-gas content.
Moreover, our non panoramic view of the Milky Way, with our solar system
deeply embedded in it, does not allow us to determine firmly important
properties of our host galaxy, such as the global SFR and its variations
across the galactic disk. High sensitivity and resolution in external
galaxies are desirable in order to link properties integrated over the disk
to the distribution of individual star-forming regions, to the
ISM components and to large scale perturbations in the disk.
Data from recent space missions, such as the {\it Galaxy Evolution Explorer
(GALEX)} and \spitzer\ satellite, allow us today to investigate these 
properties in the nearest galaxies,
before the next generation of telescopes will resolve star-forming sites
in more distant ones.

The Local Group galaxy \mm\ has a high SFR per unit area compared to
M\,31 and a low extinction towards SF regions, owing to its moderate gas content 
\citep{2003MNRAS.342..199C} 
and its low inclination \citep[54\degr,][]{2006ApJ...647L..25M}.
It is a rather quiescent blue disk galaxy,
showing no sign of recent mergers or interaction.
At a distance $D$ of only
840~kpc \citep[see][]{1991ApJ...372..455F} it allows us to examine its disk with
detailed resolution and  constitutes a local prototype for studying the ISM and
its relation to SF in blue, low luminosity objects.
Observations of the molecular, atomic and ionised gas in \mm\ have been carried out
in the past few years
\citep[e.g.][]{2000MNRAS.311..441C, 2003MNRAS.342..199C, 2003ApJS..149..343E, 2004ApJ...602..723H}
together with metallicity surveys \citep{2007A&A...470..843M,2007A&A...470..865M,
2008ApJ...675.1213R}
of the young and old stellar populations across its disk.The galaxy
has about 10\% of the total gas content in molecular form and a very
shallow metallicity gradient from 0.5 kpc to the edge of the star-forming disk.

In the first paper of this series, dedicated to the SF in \mm\
\citep[][hereafter Paper I]{2007A&A...476.1161V}, we 
used \spitzer\ maps to select a large sample of infrared (IR) emitting
sources in the 24\mi\ map of this galaxy and to identify
their counterparts in all IRAC wavebands and in \Ha. 
There is large spread in the correlation between the \Ha\ and the IR fluxes
of star-forming sites, similar to what has been found
by \spitzer\ in the faint, metal deficient local dwarf NGC\,6822
\citep{2006ApJ...652.1170C}. 
In selected
areas of this galaxy the \Ha-to-IR ratios show variations as high as a
factor of 10, contrary to the tight correlation found in the 
bright galaxy, M\,51 
\citep{2005ApJ...633..871C,2007ApJ...671..333K}.
In a forthcoming paper (Corbelli et al. 2008, in preparation)
we will use multiwavelength photometry to
characterise how stars form in individual 
star-forming sites in \mm.

In this paper, the second of this series,
we focus on the integrated IR emission and on its
radial variations to establish the role of discrete star-forming
regions in powering the dust emission across the disk.
Indeed, there has been a long debate on the role of localised warm sources
to power the dust emission and the IR fluxes in general.
Comparing IRAS and \Ha\ observations of \mm, \citet{1990ApJ...358..418R}
and \citet{1997AJ....113..236D} found that more than 70\%
of the far infrared (FIR) is linked to massive stars complexes.
On the other hand, based on ISO observations, \citet{2003A&A...407..137H} suggested
that a substantial part of the FIR emission could be due to the cold component heated
by the diffuse interstellar radiation field (ISRF).
Part of the difficulties of previous surveys in determining the grain heating source
had to do with the difficulties in resolving sources from IRAS and ISO observations.
With \spitzer, we can now
probe very faint \hii\ regions and tackle this problem more effectively.
In Paper I we have shown that discrete sources account for less
than half of the IR emission at 8 and 24\mi. Understanding the origin
of these high diffuse fractions in \mm\ will be one of the
goals of this paper. We will examine dust heating sources and the
contribution of dust in
different environments (diffuse ISM, \hii\ regions,
evolved stars) to the IR emission at several wavelengths as we move from
the center to the outermost regions of \mm.
We then use \mm\ as a prototype to test the reliability of
UV continuum fluxes, recombination lines and IR continuum fluxes
\citep[see, e.g., ][for a complete review]{1998ARA&A..36..189K}, 
as tracers of SF to possibly understand more distant galaxies, similar
to \mm\ but unresolved by present instruments.

The plan of the paper is the following: in
Sect.~\ref{sec:data} we present the data, in
Sect.~\ref{sec:radial} the integrated
luminosities in UV, \Ha\ and IR and their radial profiles, 
as well as the atomic and molecular gas distributions.
Optical and UV extinction and dust abundance are presented in Sect.~\ref{sec:gasDust}.
In Sect.~\ref{sec:sources} we discuss the origin of the IR emission
in discrete sources and in the more diffuse component, with particular emphasis
to the 24\mi\ band. In
Sect.~\ref{sec:sfr} we discuss different tracers of SFRs and
Sect.~\ref{sec:conclusion} summarises our main results.

\section{Observations and data sets} \label{sec:data}

\subsection{Ultraviolet data}

To investigate the UV continuum emission of \mm, we make use of observations by the
{\it Galaxy Evolution Explorer (GALEX)} mission \citep{2005ApJ...619L...1M}.
We use the data distributed by \citet{2007ApJS..173..185G}.
A description of \galex\ observations of \mm\ in the far--UV (FUV, 1350--1750~\AA) 
and near--UV (NUV, 1750--2750~\AA)
can be found in \citet{2005ApJ...619L..67T}, 
together with the relevant information on data reduction and calibration procedures.

\subsection{\Ha\ data}

We use the narrow-band \Ha\ image of \mm\ obtained by \citet{1998PhDT........16G}.
The reduction procedure, using standard IRAF\footnote{IRAF is distributed by the
National Optical Astronomy Observatories,
which are operated by the Association of Universities for Research
in Astronomy, Inc., under cooperative agreement with the National
Science Foundation.} tasks to subtract the continuum emission,
is described in detail in \citet{2000ApJ...541..597H}.
The total field of view of the image is $1.75 \times 1.75$~deg$^2$ ($2048 \times 2048$
pixels of 2\farcs0).

\subsection{Infrared data}

The MIR and FIR data of \mm\ are those provided by the \spitzer\ InfraRed Array Camera (IRAC) and
Multiband
Imaging Photometer for \spitzer\ (MIPS) instruments \citep{2004ApJS..154....1W,
2004ApJS..154...10F,
2004ApJS..154...25R}. The complete set of IRAC (3.6, 4.5, 5.8, and 8.0\mi) and MIPS (24, 70, and
160\mi)
images of \mm\ is described in \citetalias{2007A&A...476.1161V}: the {\it Mopex} software
\citep{2005PASP..117.1113M} was used to gather and reduce the Basic Calibrated Data (BCD). 
The complete field of view observed by
\spitzer\ is very large and allows us to achieve a complete view of
the star-forming disk of \mm, in spite of its relatively large extension on the sky.
The images were background subtracted, as explained in \citetalias{2007A&A...476.1161V}.
The spatial resolution measured on the images are 2\farcs5, 2\farcs9, 3\farcs0, 3\farcs0, 6$''$,
16$''$, and 40$''$ for IRAC 3.6, 4.5, 5.8, 8.0\mi\ and MIPS 24, 70, and 160\mi, respectively. 
The absolute photometric calibration uncertainties are better than 10\% for all IRAC \citep{2004ApJS..154...10F,2005PASP..117..978R} and the 
MIPS 24\mi\ bands \citep{2007PASP..119..994E} and are better than 20\% 
for the MIPS 70 and 160\mi\ channels \citep{2004ApJS..154...25R}.

\subsection{21-cm and millimeter data}

The radio and mm databases to infer the atomic and molecular gas distributions are the following:
Westerbork Radio Synthesis Telescope (WRST) array
\citep[$24''\times 48''$ spatial resolution]{1987A&AS...67..509D} and
Arecibo single dish \citep[$4'$ sp. res.]{1997ApJ...479..244C}
21-cm line data for the atomic gas; Berkeley Illinois Maryland Association
(BIMA) array \citep[$13''$ sp. res.]{2003ApJS..149..343E} and Five College Radio Astronomy
Observatory (FCRAO) single dish \citep[$45''$ sp. res.]{2003MNRAS.342..199C} for molecular
gas emission as traced by the CO~J=1-0 rotational line.

\section{Integrated emission and radial profiles} \label{sec:radial}

The sudden drop of the \Ha\ emission observed by \citet{1989ApJ...344..685K} 
in \mm\ , and shown also here later in this Section, defines the extent of 
the star-forming disk to be 7~kpc in radius.
In this Section, we first estimate the total luminosity of 
\mm\ in UV, \Ha\ and IR wavelengths
by integrating the observed fluxes over the star-forming disk;
next we extract radial profiles of elliptically averaged fluxes at IRAC
and MIPS wavelengths and also from the FUV, NUV, and \Ha\ images.
We will use these profiles to evaluate the radial dependence of the
average extinction which we then use to derive extinction corrected
profiles and total luminosities in UV and \Ha.
The IR emission (IRAC+MIPS) will be considered
optically thin: luminosities are not corrected, and face-on values for 
the surface brightness are given by correcting only 
for the disk inclination to the line of sight (54\degr).

\subsection{Integrated luminosities} \label{subsec:radialUV}

\begin{figure}
\includegraphics[width=\columnwidth]{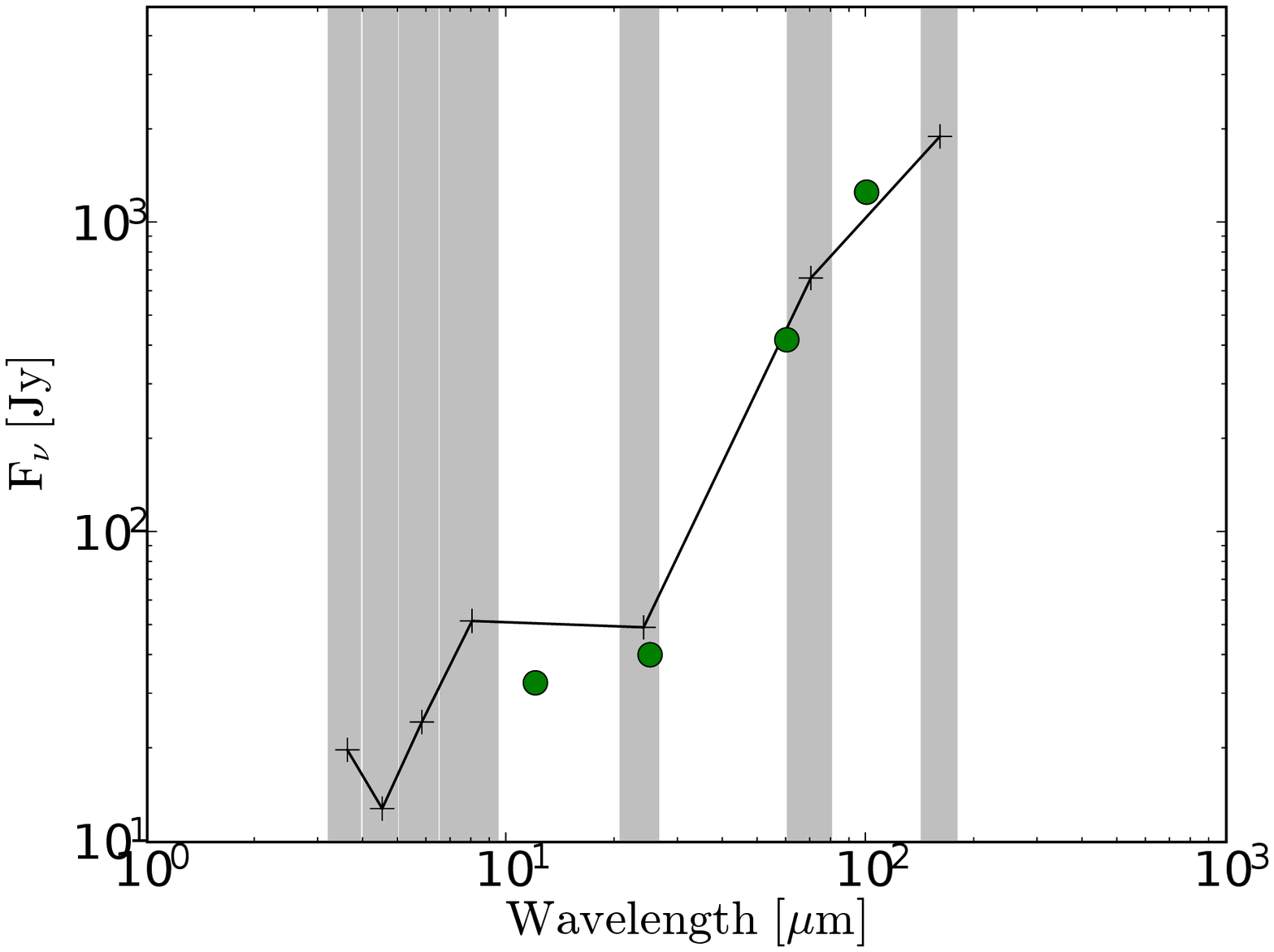}
\includegraphics[width=\columnwidth]{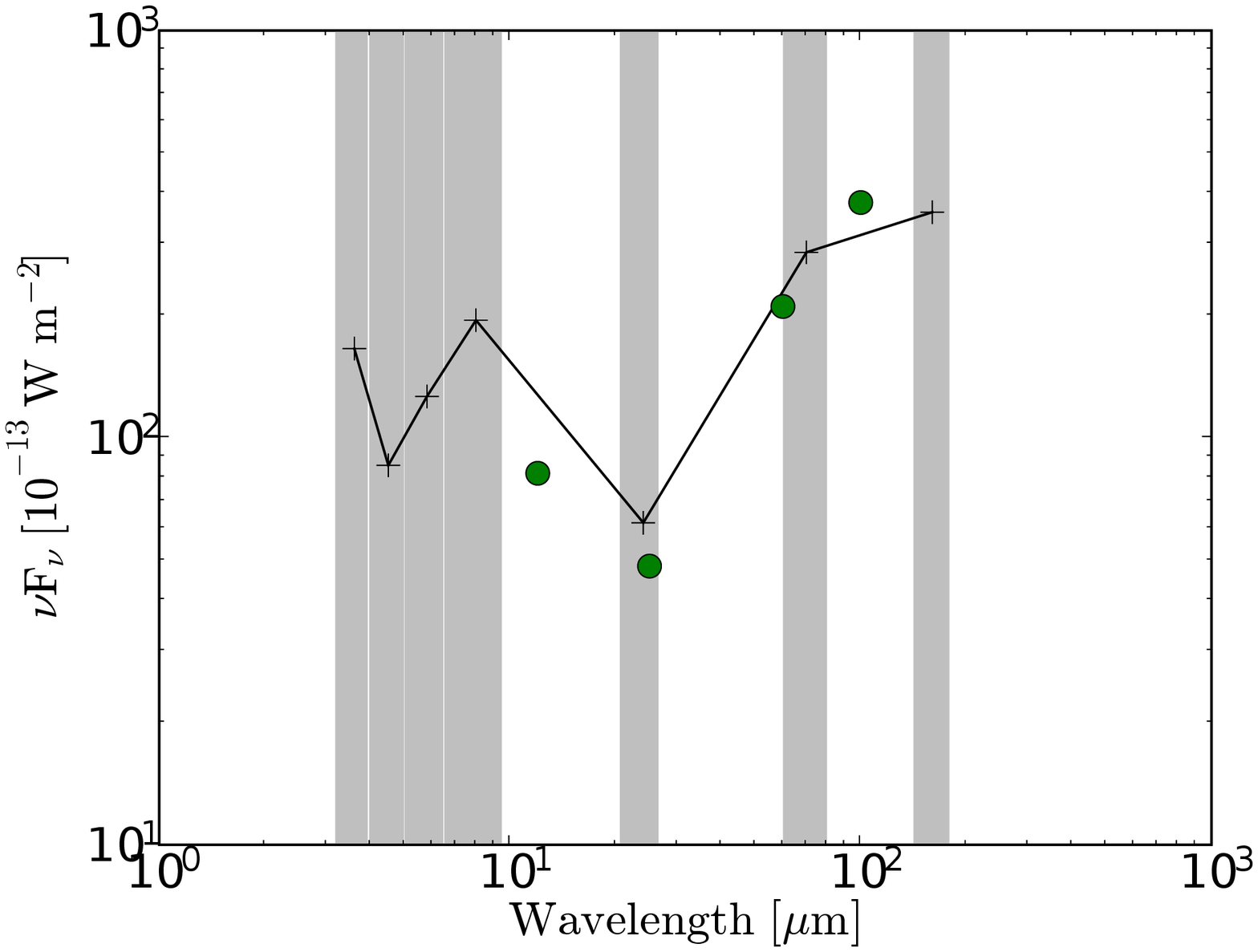}
\caption{The infrared spectral energy distribution of \mm\ is depicted
by (black) pluses (IRAC and MIPS \spitzer\ data).
The IRAS data (12, 25, 60, and 100\mi, green circles) from
\citet{1990ApJ...358..418R} are
compatible with our data within the quoted uncertainties.}
\label{fig:sedM33-v1.eps}
\end{figure}

The IR luminosities within 7~kpc
for the different IRAC and MIPS
bandpasses\footnote{Following \citet{2005ApJ...633..871C},
the convention $F,I,L$(band)~=~$\nu F_\nu, \nu I_\nu, \nu L_\nu$(band) is adopted.} 
are given in Table~\ref{tab:totFluxes}, 
together with the 3-1000\mi\ total-infrared luminosity (TIR) derived
according to the \citet{2002ApJ...576..159D} relation:

\begin{equation} \label{eq:TIR}
F({\rm TIR}) = 10^{-14} \times (19.5\ F_\nu(24) + 3.3\ F_\nu(70) + 2.6\ F_\nu(160)) ,
\end{equation}

\noindent
with $F$(TIR) in W~m$^{-2}$ and $F_\nu$ in Jy. 
The TIR luminosity of \mm\ is $1.8\times 10^9$~L$_\odot$.
The TIR luminosity we estimated from 8 and 24\mi\ in
\citetalias{2007A&A...476.1161V}  
was $1.0 \times 10^9$~L$_\odot$, but within a shorter radius (5~kpc), 
hence the two
values are quite consistent.
Using the values in Table~\ref{tab:totFluxes}, we show in Fig.~\ref{fig:sedM33-v1.eps} the
mid- and far-IR spectral energy distribution (SED) together with
IRAS data from \citet{1990ApJ...358..418R} which, given the uncertainties,
turn out to be in satisfactory agreement with our measurements.
In the lower panel of Fig.~\ref{fig:sedM33-v1.eps}, the SED is
depicted as $\nu I_\nu$ 
which can be directly compared with the IR SED of 71 nearby galaxies 
by \citet{2005ApJ...633..857D}.
The SED results from two major components: the stellar photospheric emission,
which decreases from 3.6 to longer wavelengths,
and the rising continuum due to dust emission.
The peak at 8.0\mi\ gives clear evidence of the presence of PAHs in \mm.

The integrated FUV and NUV luminosities within 7~kpc
are $3.9 \times 10^{42}$~erg~s$^{-1}$ and 
$3.2 \times 10^{42}$~erg~s$^{-1}$ respectively
\citep{2007ApJS..173..185G}.
The \Ha\ luminosity \citep[as given by][]{2000ApJ...541..597H} is
$3.1 \times 10^{40}$~erg~s$^{-1}$.
Again, Table~\ref{tab:totFluxes} summarises the total fluxes, and 
luminosities measured
within the SF disk of \mm\ before applying any extinction
correction, and with extinction corrections derived in 
Sect.~\ref{subsec:extin}.

\begin{table}
\begin{center}
\caption{Total fluxes and luminosities within 7~kpc. \Ha, FUV and NUV
luminosities corrected for extinction are also shown.} \label{tab:totFluxes}
\begin{tabular}{c c c c}
\hline \hline
Wavelength & Total fluxes & L & L(corr) \\

[\mi] & [Jy] & [erg s$^{-1}$] & [erg s$^{-1}$]\\
\hline
3.6 & 19.9  & 1.4 10$^{42}$& \\
4.5 & 12.8  & 7.2 10$^{41}$& \\
5.8 & 24.4  & 1.1 10$^{42}$& \\
8.0 & 51.8  & 1.6 10$^{42}$& \\
24  & 49.4  & 5.2 10$^{41}$& \\
70  & 666.  & 2.4 10$^{42}$& \\
160 & 1909. & 3.0 10$^{42}$& \\
TIR &       & 6.8 10$^{42}$& \\
\Ha &       & 3.1 10$^{40}$& 4.2 10$^{40}$\\
FUV &       & 3.9 10$^{42}$& 7.0 10$^{42}$\\
NUV &       & 3.2 10$^{42}$& 6.3 10$^{42}$\\
\hline
\end{tabular}
\end{center}
\end{table}

\subsection{IR radial profiles}

Previous work has suggested that the SF disk in \mm\ shows three
morphologically distinct areas: 
first, the innermost 1~kpc where a weak bar is in place and
metallicity is enhanced; 
second, the inner 3.5~kpc disk where
spiral arms and GMCs are prominent; 
and finally the outer disk where structures weaken 
before the edge of the SF disk occurs at about 7~kpc.
It is therefore of interest to examine how the old and young stellar populations and the
dust emission vary among these three regions and in general as a function of radius.

To extract the IR radial profiles from images, we used the IRAF task {\it ellipse},
fixing the center to the nominal position of \mm\ (01$^{\rm h}$33$^{\rm m}$50\fs90,
+30\degr39\arcmin35\farcs8), the axial ratio to 0.59 as for a fixed disk inclination of
54\degr\, and the position angle to 22.5\degr.

\begin{figure}
\includegraphics[width=\columnwidth]{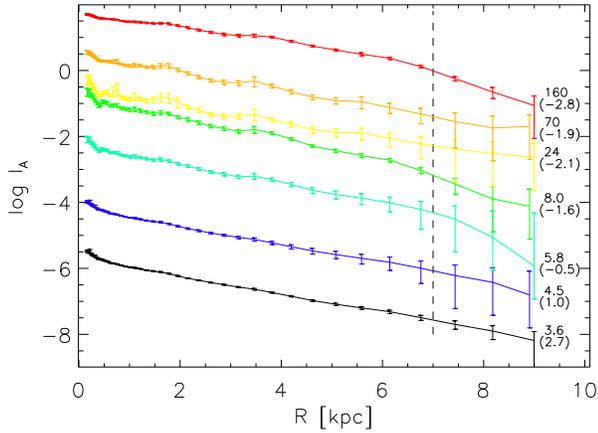}
\caption{Radial profiles, using logarithmical bins, of the \spitzer\ MIPS+IRAC surface brightness in arbitrary
units. For clarity, results for the different bandpasses have been shifted in the 
vertical direction (within brackets, the logarithmic shift to be added to the ordinates 
to obtain the real surface brightness in mJy~pc$^{-2}$).
The dashed vertical line at 7~kpc marks the edge of the
star-forming disk, see text.}
\label{fig:profilesIR}
\end{figure}

We show in Fig.~\ref{fig:profilesIR} the elliptically averaged surface brightness
(I$_A$) in the IRAC and MIPS bands, as a function of galactocentric radius 
after correction flux densities to face on values.
In Fig.~\ref{fig:profilesIR}, radial bins increase logarithmically:  
the multiplicative factor between two
consecutive elliptical semiaxes is 1.1, that is an increment of 10\%.
The errors are computed by quadratically adding the error
of the mean along the ellipse provided by {\it ellipse} task itself
and the error on the subtracted background.
This background error is estimated for each image by considering
the noise and flatness of the image in the most peripheral areas
(considering 30 different areas around the borders).
These estimates of the background uncertainty
agree pretty well with the level of oscillation of the mean
intensity from {\it ellipse} when it achieves background flux levels.

The extraction of the IRAC and MIPS-24 profiles was performed on images where 
foreground stars had been previously removed; we observed, in fact, that they introduce
significant local enhancement in the profiles, particularly at 3.6 and 4.5\mi, 
the most sensitive to photospheric emission.
At all wavelengths, the average surface brightness shows an enhancement in the innermost 
0.5~kpc region (see Sect. \ref{uvha:prof}), 
then declines outwards exponentially. 
While the old stellar population dominates at short wavelengths,
PAHs and Very Small Grains (VSGs) in the surrounding of hot stars
are expected to contribute at 8 and 24\mi\ respectively. 
Hence these last profiles display more features, 
especially in the inner disk where spiral arms are prominent. 
These are visible also in the emission at 70\mi, while
at 160\mi\ the coarse resolution limits
the detection of such features in the elliptical averages.

The $\chi^2$ fit exponential scale lengths fitted from 0.5\,kpc to the edge of the
star-forming disk are reported in Table~\ref{tab:scales}.
The scale lengths are computed after rebbining the data into 
radial bins of equal size (50 pixels i.e. 0.24~kpc) taking into  
account the relative brightness uncertainties).
All MIPS scale lengths are, within 3~$\sigma$, compatible 
with those determined by \citet{2007A&A...472..785T} (see their Table~7).
The profiles at 3.6, 4.5, 24, and 70\mi\ are consistent
with a single exponential
scale length throughout the star-forming disk. This is $\sim$1.55~kpc for
the old stellar population emission and
$\sim$1.75~kpc for the dust emission at 24 and 70\mi.
Two different slopes for the inner (0.5--3.5~kpc) and outer (3.5--7.0~kpc) star-forming 
disk are found instead at 5.8, 8.0 and 160\mi. At these wavelengths the IR brightness
declines more steeply beyond 3.5~kpc and drops at the edge of the star-forming disk.
Drops at the disk edge are less evident at other wavelengths, even though the
 uncertainties at large radii make the data consistent either with the
no-drop hypothesis or with the one that undoubtedly exists, namely at 160\mi.
We also note that the disk of \mm\ starts warping beyond 7~kpc, which adds
uncertainties which have not been taken into account.

\begin{table}
\begin{center}
\caption{Scale lengths in kpc for the radially declining
surface brightness at various IR wavelengths.}
\label{tab:scales}
\begin{tabular}{c c c c}
\hline \hline
Wavelength & 0.5 : 3.5~kpc & 3.5 : 7~kpc & 0.5 : 7~kpc \\
\hline
3.6\mi\ & 1.51 $\pm$ 0.02 & 1.61 $\pm$ 0.05 & 1.56 $\pm$ 0.01\\
4.5\mi\ & 1.52 $\pm$ 0.04 & 1.60 $\pm$ 0.15 & 1.55 $\pm$ 0.03\\
5.8\mi\ & 1.59 $\pm$ 0.06 & 1.36 $\pm$ 0.13 & 1.61 $\pm$ 0.04\\
8.0\mi\ & 1.52 $\pm$ 0.07 & 1.21 $\pm$ 0.05 & 1.44 $\pm$ 0.02\\
24\mi\  & 1.71 $\pm$ 0.10 & 1.59 $\pm$ 0.20 & 1.77 $\pm$ 0.06\\
70\mi\  & 1.78 $\pm$ 0.10 & 1.55 $\pm$ 0.14 & 1.74 $\pm$ 0.05\\
160\mi\ & 2.41 $\pm$ 0.09 & 1.48 $\pm$ 0.04 & 1.99 $\pm$ 0.02\\
\hline
\end{tabular}
\end{center}
\end{table}

The similarity of the 5.8, 8.0, and 160\mi\ emission may be understood
by comparing radial trends of PAH emission and cold dust.
PAHs contribute to the 5.8\mi\ emission and even more to the 8.0\mi\ emission band.
\citet{2002A&A...385L..23H} and 
\citet{2008arXiv0806.2758B} have shown that PAH and
cold dust emission correlate spatially and inferred that the 
diffuse ISRF is the power source of both emissions.
Indeed, \mm\ both 8\mi\ and 160\mi\ show a steeper decline in the outer disk.
This might be due to a decrease in the surface density of the 8 and 160\mi\ carriers
since the decrease of the energy density of the ISRF in the outer disk is not steeper
(see also Sec.~\ref{uvha:prof}).
However, the 8\mi\ profile declines with a steeper slope throughout the whole disk
implying that the 8 to 160\mi\ ratio also declines
as we move outwards.
One possibility is that the PAH abundance or emission is more sensitive to quantities
which vary across the disk, such as the metallicity, the efficiency of supernovae shocks
in destroying the molecules \citep{2006ApJ...641..795O}, or the disk gravity, which
affects the cooling and ionisation balance in the ISM (see Sect. \ref{hih2:prof}).
\mm\ has a shallow metallicity gradient, and the oxygen abundance [O/H] is always
above -3.8, except at the edge of the star-forming disk \citep{2007A&A...470..865M}.
This casts some doubt that the metallicity is the main driver of the fast radial
decline of PAH emission.

The scale length inferred for the old stellar population using the 3.6 and 4.5\mi\
emission is consistent with what earlier studies have derived in the inner part of
the disk using photometry in the K-band
\citep[1.4$\pm0.1$~kpc,][]{1994ApJ...434..536R}. Ground based near IR data
for this galaxy does not extend radially outwards as far as \spitzer\ data, but a
colour map of the 3.6-2.2\mi\ emission using the
2MASS (J+H+K) image \citep{2003AJ....125..525J} reveals a radial gradient
between 0.5 and 5~kpc due to the smaller scale length of the J+H+K
emission (of order 1~kpc).
This might imply that intermediate age, cool supergiants
contribute a substantial fraction of the NIR emission. 
The colour map follows the spiral
morphology, suggesting that cool supergiants are more likely candidates than giants.

\subsection{UV and \Ha\ line radial profiles} \label{uvha:prof}

\begin{figure}
\includegraphics[width=\columnwidth]{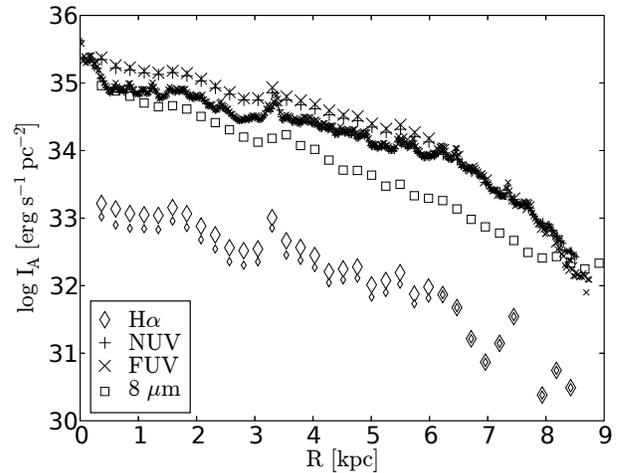}
\caption{Brightness radial profiles  in the FUV (small crosses), NUV (small pluses)
and \Ha\ (small diamonds). Extinction corrected brightness profiles in
bins of 0.24~kpc are also shown in the FUV (large crosses), NUV (large pluses) 
and \Ha\ (large diamonds). 
I$_A$ is here converted to an absolute surface luminosity scale,
that is multiplied by $\nu \times 4 \pi D^2$.
The 8\mi\ radial profile (squares) is plotted for
comparison with the UV brightness.}
\label{fig:profilesUVHa}
\end{figure}

To evaluate the recent SFR across \mm\  we will use the FUV and \Ha\
radial profiles. It is more reliable to compute the SFR from integrated emission
across the disk rather than from single SF sites. This is because SF sites lose
part of their energetic photons in the surrounding ISM and the stars 
themselves diffuse out of their birthplace over time.
FUV is a diagnostic of a somewhat younger (30-100~Myr) star formation than NUV (100-300~Myr),
while \Ha\ relates to the youngest stellar population.
In Fig.~\ref{fig:profilesUVHa}, we show the observed radial profiles of the NUV, FUV and \Ha\
emission \citep{2005ApJ...619L..67T} in bins of 0.24~kpc. 
\Ha, FUV and NUV display very similar radial profiles
which points to a scarce radial evolution of the SFR in the last 300~Myr.
Unlike other galaxies \citep{2007ApJS..173..538T},
in \mm, the UV emission does not extend farther than \Ha.
In addition, we observe no strong gradient in the colour ratio FUV$-$NUV;
\mm\  is only slightly bluer with increasing radius opposite to
the more pronounced trends found by \citet{2005ApJ...619L..71B} in M\,51 and M\,101.
This confirms the lack of an appreciable age gradient
in the young stellar populations, and precludes 
strong radial variations of the UV spectrum.

UV and \Ha\ show a strong enhancement in the 
innermost 0.5~kpc likely due to the effect of
the bar fueling of the circumnuclear region \citep{2007ApJ...669..315C}
and that is another reason why we excluded that region from our global
fits. The radial scale lengths of the UV and \Ha\
brightness are shown in the fourth column of Table 3. They do not have
significant variations between the inner and the outer disk and hence we
display only their average values across the whole disk.
In the last column of Table 3
we give these value after correcting for extinction, as described in
the next Section. 

\begin{table}
\begin{center}
\caption{Scale lengths for the H$\alpha$, UV (uncorrected and corrected for extinction),
the molecular and total gas emission.} \label{tab:scales2}
\begin{tabular}{c c c c c}
\hline \hline
Line & Wavelength & Range & $h_r$ & $h_r$(corr) \\
 & & kpc & kpc & kpc\\
\hline
\Ha\ & 6563 \AA & [ 0.5 : 6.0 ] & 2.0$\pm$0.2  & 1.8$\pm$0.1\\
NUV & 1750--2750 \AA & [ 0.5 : 6.0 ] & 2.3$\pm$0.1 & 2.1$\pm$0.1\\
FUV & 1350--1750 \AA & [ 0.5 : 6.0 ] & 2.4$\pm$0.1 & 2.2$\pm$0.1\\
CO BIMA & & [ 0.5 : 6.0 ] & 1.4$\pm$0.1 & -- \\
CO FCRAO& & [ 0.5 : 6.0 ] & 2.4$^{+1}_{-0.4}$  & -- \\
Total gas &   & [ 0.5 : 6.0 ] & 16$^{+18}_{-5}$ & --\\
\hline
\end{tabular}
\end{center}
\end{table}

All three young SF tracers --\Ha, FUV, and NUV-- display a compatible 
scale length of about 2~kpc while the oldest stellar populations (3.6 and 4.5\mi) 
exhibit a shorter one (approximately 1.5~kpc). 
The fact that the youngest stellar component has the longest
scale length supports the inside-out disk growth scenario for \mm\ 
\citep[e.g.][]{2007A&A...470..843M}.
In general, our data suggest that dust emission at MIPS wavelengths in
\mm\ is also declining radially with longer scale lengths than the old stellar population, 
especially in the inner disk. The average dust emission scale length over the SF disk
is consistent with the one traced by the young stellar populations even though the
shorter scale length of the IR in the outer disk, especially at 160\mi\, suggests
a decline of the dust surface density.

\subsection{Atomic and molecular gas distributions}\label{hih2:prof}

\begin{figure}
\includegraphics[width=\columnwidth]{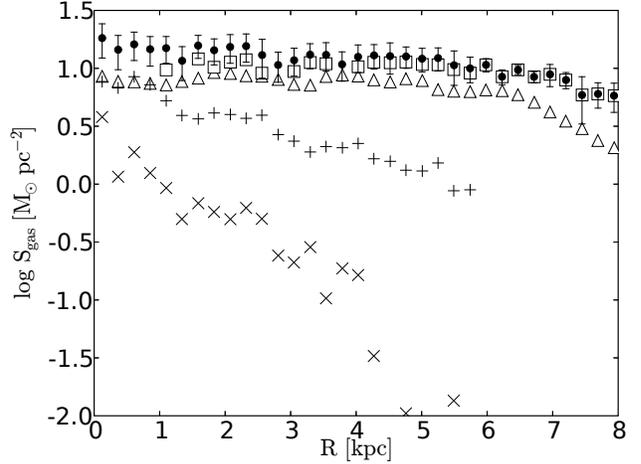}
\caption{Gas surface densities. The atomic gas data is shown by the open squares and open triangles
for the Arecibo and Westerbork data, respectively. The molecular gas is shown by crosses and pluses
for the BIMA and FCRAO data, respectively. The sum of the atomic (using the Arecibo data, 
except for the first four inner radii where we used the values from Westerbork) and molecular gas
(FCRAO) is also displayed with filled circles. Error bars show the dispersion in each bin.}
\label{fig:gas}
\end{figure}

In Fig.~\ref{fig:gas} we show the atomic and molecular gas column density
as a function of galactocentric radius and Table~3 gives their radial scale lengths.
The total gas distribution, sum of the \hi\ (using the Arecibo data, except for the
first four inner radii where we used the values from Westerbork) and of the diffuse
molecular gas (using the FCRAO data), is also shown in Fig.~\ref{fig:gas}.
Uncertainties on the radial scale lengths for the CO-FCRAO and total gas 
are computed using the $\chi^2$ statistic
taking into account the dispersion of the data in each radial bin. In Figure~\ref{fig:gas}
we display for clarity only the dispersions for the total gas surface density
distribution, which is of relevance for this paper. The CO-BIMA scale length
and its uncertainty are taken from \citet {2003ApJS..149..343E}.
The \hi\ surface density is remarkably constant with radius
\citep[as already noted by][]{1997AJ....113..236D}, with
a value of $\sim$10~M$_\odot$~pc$^{-2}$, over most of the galaxy.
Only beyond 6~kpc, a cutoff is observed in the Westerbork data,
likely due to interferometer losses where the ISM is more uniformly diffuse.
The situation is different for the molecular gas:
the CO emission decreases steeply outwards,
implying a less efficient conversion from atomic to molecular gas at larger radii.
As discussed by \citet{2004ApJ...602..723H} the molecular fraction is regulated by the
balance between formation rate, related to the hydrostatic pressure
in the galactic plane, and dissociation rate related to the ISRF.
The hydrostatic pressure decreases rapidly outwards because of the decrease of
the stellar disk surface density,
which dominates over the gas surface density inside 5~kpc \citep{2003MNRAS.342..199C}.

\section{Dust in the disk} \label{sec:gasDust}

\subsection{UV and Optical extinctions} \label{subsec:extin}

To correct the UV and \Ha\ luminosities for
internal extinction we adopt the prescription by \citet{2001PASP..113.1449C} which
relates \av\ to the TIR and FUV luminosities :

\begin{equation} \label{eq:avFUV}
A_{\rm V} = C \times 1.76 \times \log\left(\frac{1}{1.68} \times \frac{L(\rm TIR)}{L(\rm FUV)} +
1\right) \quad .
\end{equation}

\noindent
This gives an estimate of the average extinction of radiation from
sources within the \mm\ disk.
The value of $C$ depends on whether we consider extinction of
the stellar continuum or of the radiation from the ionised gas.
In the model considered by \citet{2001PASP..113.1449C}, the dust is 
driven to the outer part of star-forming regions by first-generation 
massive stellar winds and supernova events.
The dust will then produce a screen-like extinction
for the radiation from central stars and from the ionised gas.
Previous generations of stars,
still contributing substantially to the UV emission,
will instead be interspersed within the dust,
since they tend to diffuse out from the central region.
Consequently, the extinction of the stellar continuum from older stars  will be
lower than the  extinction suffered by younger stars and by the ionised gas. 
In practice, $C=1$ for the
ionised gas and $C=0.44$ for the stellar continuum.
In the case of \mm\ this scenario is supported by the distinct morphology of the
\galex\ and \Ha\ maps: the \hii\ regions are very localised,
the UV much less so. However since about half of the H$\alpha$ radiation 
in \mm\ is diffuse we shall consider an average value: $C=0.7$ for the ionised gas.
This model assumes that the bulk of dust heating, which is re-radiated in the IR, 
comes from massive stars, the main contributors of UV photons.
As we will see later in this paper, this assumption might not be valid for the
MIR radiation at 24\mi\, whose contribution to the TIR, however, is small, about 10\%.

Using a standard extinction curve, \av~=~$3.1\,E(B-V)$, 
and following \citet{2005ApJ...619L..55S},
i.e. $A_{\rm FUV} = 8.29 \,E(B-V)$, we can infer $A_{\rm FUV}$. 
Globally, \av\ is found to be 0.30~mag (before applying face-on corrections)
and hence $A_{\rm FUV}=0.53$~mag,
consistent with the findings of \citet{1996A&A...306...61B}
which give an upper limit of 0.7~mag;
hence the extinction corrected FUV luminosity of
\mm\ is $6.3 \times 10^{42}$~erg~s$^{-1}$.
Using $A_{\rm NUV} = 8.18 \times E(B-V)$ \citep{2005ApJ...619L..55S},
$A_{\rm NUV}=0.52$~mag on average
and the corrected NUV luminosity is $5.5~\times~10^{42}$~erg~s$^{-1}$.

\begin{figure}
\includegraphics[width=\columnwidth]{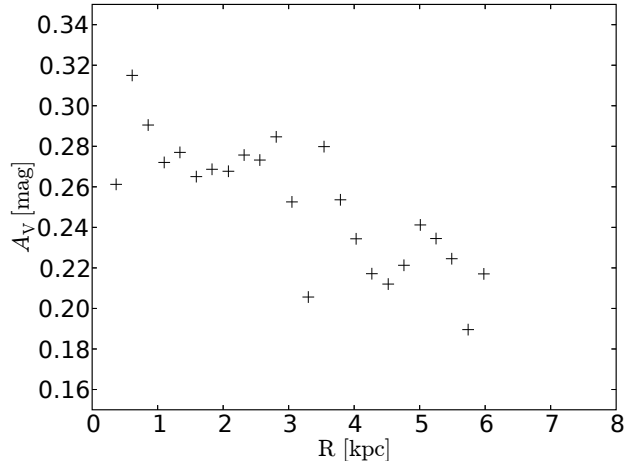}
\caption{\av\ face-on values for the stellar continuum averaged in radial 
bins of 0.24~kpc, as a function
of galactocentric radius.}
\label{fig:av}
\end{figure}

In Fig.~\ref{fig:av}, we show 
the average  face-on \av\ value for the stellar continuum 
as a function of radius. \av\ displays a
rather constant value of about 0.27~mag 
up to 3~kpc, than drops gently
to below 0.22~mag  beyond 6~kpc.
The drop of the UV luminosity at larger radii, 
due to the effective radial cut-off of the star
forming disk, 
does not allow tracing the extinction further out, 
where in fact it may be considered negligible. 

Extinction is found to be globally rather low, in agreement with
\citet{2007A&A...475..133T}.
Nevertheless,
extinction can be enhanced locally: in the cores of \hii\ regions
values of \av~$\sim$~1~mag are common \citep{1995AJ....110.2715M, 1997AJ....113..236D}.
The optical extinction,
measured from the \Ha\ to H$\beta$ ratio in selected \hii\ regions,
is found to have a large
scatter with an average of 0.65~mag and with no clear radial dependence
\citep{2007A&A...470..865M}.
Since, as noted above, the gas in \hii\ regions should in principle 
suffer from a higher
extinction than field stars, our average value of 0.25 mag for the
stellar continuum translates into \av\ $\simeq 0.57$ mag for the
\hii\ regions, very close to the finding of \citet{2007A&A...470..865M}.

From the radial profile of $A_{\rm V}$, we can recover the UV and \Ha\
radial profiles corrected for extinction. These are shown in Fig.~\ref{fig:profilesUVHa}
and the relative scale lengths are given in Table~\ref{tab:scales2}
(FUV~=~2.2~kpc~$\pm$~0.1; NUV~=~2.1~kpc~$\pm$~0.1; \Ha~=~1.8~kpc~$\pm$~0.1).
As it can be seen, extinction corrections are small and do not dramatically affect 
the profile slope. Upon correction for extinction at each radial bin,
we find the total FUV luminosity to be $7.0 \times 10^{42}$~erg~s$^{-1}$.
We deem this value more
accurate than the $6.3 \times 10^{42}$~erg~s$^{-1}$ obtained above using
the value of $A_{\rm FUV}$ averaged over the whole disk.
Similarly, the total NUV luminosity radially corrected for extinction is
$6.3 \times 10^{42}$~erg~s$^{-1}$.
Following \citet{2001PASP..113.1449C} we retain a global
estimate of extinction for the ionized gas of 0.33~mag at \Ha\ wavelength
($C=0.7$). The corrected \Ha\ luminosity is $4.2~\times~10^{40}$~erg~s$^{-1}$.

\subsection{Dust-to-gas ratio}

Since \av\ is proportional to the dust column density,
its ratio to the gas column density can be used to infer how
the average dust-to-gas ratio varies across the \mm\ disk.
As discussed above, towards \hii\ regions extinction is higher than
what Fig.~\ref{fig:av} shows.
Thus we expect the dust column density to be larger than
what can be inferred from Figure~\ref{fig:av}.
There is also the possibility of an additional dust component 
dispersed in the ISM and not heated by the young stellar
population which has not been taken into account in the computation
of \av.

In order to derive the dust column density
from emission processes we shall derive the dust optical depth
using the 160\mi\ brightness and the dust temperature (see Section~\ref{sub:dustTemp}
for details on the dust temperature). The method used and 
the derived radial distribution of $\tau_{160}$, the optical depth at 160\mi\, 
are very similar to those reported by  
\citet{2007A&A...472..785T} for the same galaxy.
The opacity decreases by a factor 10 from the center to the edge
of the star-forming disk. We then convert the dust opacity to
the global visual extinction through half of the face-on disk of \mm\ using:

\begin{equation}\label{eq:avtau}
A_{\rm V}={1.086\over 2} \times 2500\times \tau_{160} \quad .
\end{equation}

\begin{figure}
\includegraphics[width=\columnwidth]{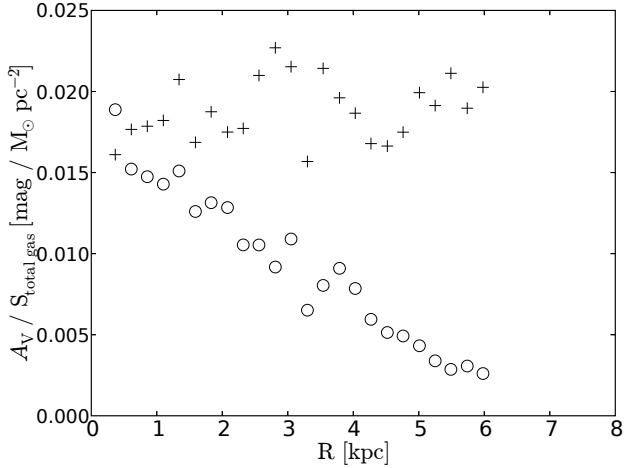}
\caption{The visual extinction normalized to the total gas column density
as a function of the galactocentric radius. Plus symbols show this ratio
for \av\ derived from Eq.~\ref{eq:avFUV}, open circles for \av\ as from
Eq.~\ref{eq:avtau}.
}
\label{fig:dustGas}
\end{figure}

\noindent Fig.~\ref{fig:dustGas} shows the ratio of \av\ to the total gas surface density
as a function of radius using the \av\ values derived both from Eq.~\ref{eq:avFUV}
and from Eq.~\ref{eq:avtau}. 
The gas is the sum of the \hi\ (from the Arecibo data, except the
first point which is from Westerbork) and molecular gas (from the FCRAO 
data)\footnote{If
only interferometric data were used to estimate the \hi\ density,
due to the lower sensitivity and incomplete low spatial-frequency sampling, the
dust-to-gas ratio would instead
rise sharply in the outer part of the galaxy.}.
There is a clear discrepancy: going radially outwards
the dust-to-gas ratio derived from UV absorption does not show any 
clear dependence on radius. The dust-to-gas ratio derived from the IR emission
decreases steadily. Clearly both methods
have a caveat: the use of Eq.~\ref{eq:avFUV} is valid under the assumption that
all dust is heated by UV radiation, in the converse case it overestimates \av.
Also it depends on geometrical factors such as the clustering of hot stars,
which emit the UV radiation.  For the computation of $\tau_{160}$
we have assumed a single dust component in each radial bin (according to the temperatures shown in Fig.~\ref{fig:dustTemp}):
in the presence of several temperature components the
characteristic temperature one derives is biased to the higher values, 
resulting in an underestimate of the dust optical depth. 
Moreover the conversion factor between the optical depth in the IR and the one in the visual is highly uncertain:
values range from 2000 to 5000 for our Galaxy \citep{1981ApJ...246..416W,1984ApJ...285...89D}. The dust composition and size distribution in \mm\ can also be different from that in our Galaxy and make precise calibration difficult \citep[small particules and highly reflective materials will tend to increase the value of the conversion factor, e.g.][]{1993Icar..106...20M}.
Given these uncertainties, any conclusion about the average radial variations of the dust-to-gas ratio 
and of the dust optical depth in \mm\ would be highly speculative.

\begin{figure}
\includegraphics[width=\columnwidth]{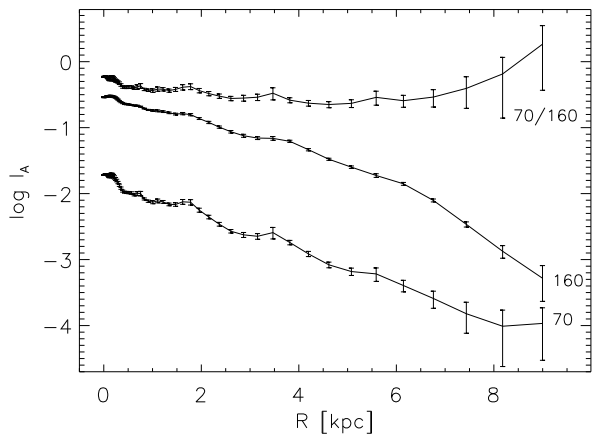}
\caption{Elliptically averaged 70 and 160\mi\ surface brightness and their ratio as a 
function of the galactocentric radius, following logarithmical bins. The two brightness profiles are in arbitrary
units and vertically displaced for clarity.}
\label{fig:70_160}
\end{figure}

\begin{figure}
\includegraphics[width=\columnwidth]{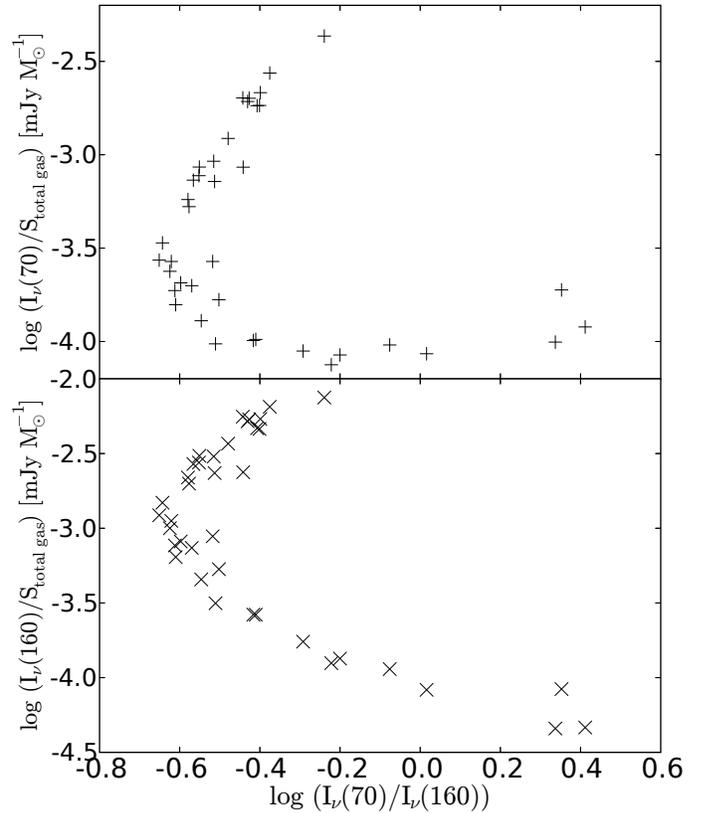}
\caption{Elliptically averaged FIR emission per unit gas versus FIR colour. They are displayed
for the 70\mi\ emission ({\it upper panel}) and for the 160\mi\ emission ({\it bottom panel}). 
In both panels, the galactocentric radius increases from the upper to the lower points.}
\label{fig:ratio3.eps}
\end{figure}

\subsection{Dust temperature} \label{sub:dustTemp}

If the dust column density has a shallower radial decrease
than the light profiles of the stellar population which is heating the dust,
it is conceivable that the radial decrease of the 70 and 160\mi\ emission is 
due to a decrease of the dust heating radiation field rather than to a decrease of
the dust column density. 
If this picture is correct and the distribution of
grain sizes does not change radially,
any variation in the ratio of gas to 70 and 160\mi\ emission 
is likely to be linked to dust
temperature variations and cooler grains are
expected in the outer part of the galaxy.
The radial behaviour of the 70/160 colour is shown in 
Fig.~\ref{fig:70_160}.

\begin{figure}
\includegraphics[width=\columnwidth]{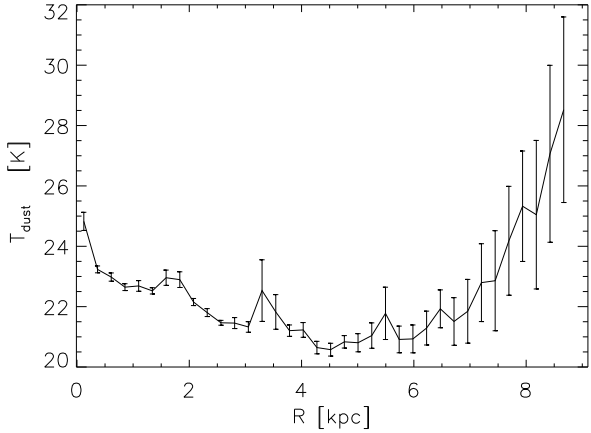}
\caption{Dust temperature as a function of radius, estimated by 
fitting a blackbody to the 70 and 160\mi\ 
elliptical emissions, in bins of 0.24~kpc.}
\label{fig:dustTemp}
\end{figure}

There are clear oscillations, in coincidence with the spiral pattern,
where the colour gets bluer due to the presence of large star-forming complexes.
Overall this ratio decreases in the inner disk, where  
hot sources contribute substantially to the FIR emission,
then remains constant up to the edge of the star-forming disk.
The ratio flattens out in the outer disk: in fact the scale length 
of the 160\mi\ emission, which was larger than that of the 70\mi\ emission
in the inner disk, decreases and emission at both wavelengths falls off
radially in the same way. At 7~kpc the data suggest an uprise but the large
error bars and the possible disk distortion due to the warp prohibit
any definitive conclusion.
By fitting an optically thin thermal spectrum 
\citep[$\kappa_\lambda \propto \lambda^{-2}$, see][ and references 
therein]{2007MNRAS.379..974H}
to the elliptical averages at 70 and 160\mi,
we derive the integrated FIR emission by integrating an optically thin 
thermal spectrum  between 40 and 1100\mi, (which is the wavelength range 
which defines the FIR).
The dust temperature, displayed in Fig.~\ref{fig:dustTemp}, decreases from 25~K in the innermost
region to 21~K at about 4~kpc, then shows a flat plateau up to 
the edge of the star-forming disk where it increases.
The integrated emission will be used in Sect.~\ref{sec:sfr}
to estimate the FIR luminosity and to derive
the conversion factor between FIR luminosity and SFR.

Figure~\ref{fig:ratio3.eps} shows the ratio of the 70 or 160\mi\ surface brightness
to the total gas column density as a function of the 70 to 160\mi\ ratio.
High values of the 70 or 160\mi\ surface brightness
to gas column density  ratio correspond to the inner disk. Here
the strength of the emission at 70 and 160\mi\ 
is correlated primarily with dust heating and not with dust abundance
since the FIR colour becomes redder as the dust
emission per gas unit decreases. 
In the outer disk this is not so evident: the dust temperature stays constant 
out to the edge of the SF disk in the presence of a radially decreasing
radiation field. If the dust column density does not change much 
and the UV radiation is the only heating source for grains,
a change in the spectral energy distribution is required.
But the lack of changes in the UV colour does not support this picture.
A change in the dust abundance, as the steep radial decline of the 160\mi\ 
emission seems to suggest, is more likely,
assuming that the dust size distribution does not vary. 
There is however some anomaly in the distribution of evolved stars
which suggests that the building up history of the \mm\  disk has been rather complex
and provides an alternative solution. An excess of Carbon Stars at large
galactocentric radii has been found
\citep[e.g.][]{2005AJ....129..729R,2007ApJ...664..850M,2008arXiv0805.1143C}.
Accretion events in the past brought some metal poor material beyond 4~kpc
thus favoring the production of Carbon Stars. 
The presence of additional radiation from those Carbon-rich giants in the outer
disk can heat the dust in their proximity thus providing an additional
contribution to the 70\mi\ surface brightness. We shall address better in the
next Section the role of evolved stars in powering the unresolved emission at
shorter wavelengths. 
Some caution is needed here in drawing any definitive conclusion because
at large radii, the 70\mi\ image has a high noise level due to residual
artifacts (see Paper I). Concerning possible changes in the grain size
distribution we cannot fully constrain it with available data 
since we are not sensitive to emission from a very cold dust component.

\section{Discrete sources and diffuse emission} \label{sec:sources}

In this Section, we derive the emission associated with discrete sources and compare this with
the total emission in order to derive diffuse fractions as a function of radius. In general, \mm\
is a galaxy with high diffuse fractions, the origin of which is still a subject of debate.
The UV diffuse fraction of \mm\ was studied by \citet{2005ApJ...619L..67T} to obtain indications
on the location and nature of the non-ionising photons. The FUV diffuse fraction appears to be
remarkably constant (equal to about 0.65) as a function of radius. Only beyond the
edge of the star-forming disk
a slight increase of the FUV diffuse fraction could indicate that some of the
emission was not produced locally but rather scattered by dust. The \Ha\ diffuse fraction
was derived by \citet{2000ApJ...541..597H} and is about 0.4 throughout the SF disk.

In \citetalias{2007A&A...476.1161V}, we used the
SExtractor software \citep{1996A&AS..117..393B} to compile a catalogue of 515 sources emitting at 24\mi.
Following the process detailed in \citetalias{2007A&A...476.1161V}, 
we can extract sources at 8\mi, and in \Ha. 
The catalogue at 8\mi\ comprises 516 sources, and the \Ha\ catalogue 413.
The coarse resolution at 70 and 160\mi\
allows the extraction of only the most prominent sources, about 150 in total;
in any case, the emission associated with discrete sources 
would be highly underestimated at these longer wavelengths and we will not discuss them further.

\begin{figure}
\includegraphics[width=\columnwidth]{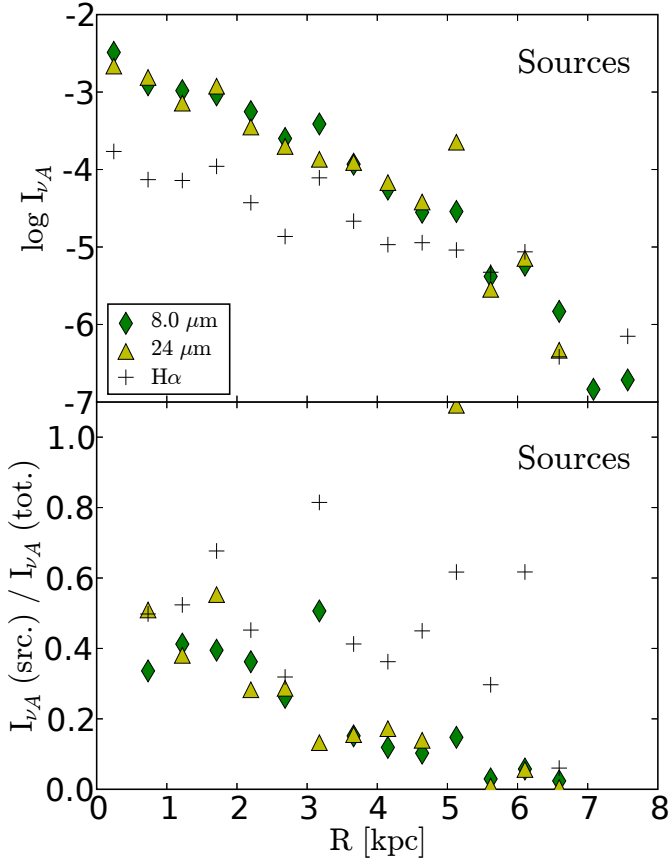}
\caption{{\it Upper panel:} Infrared (in units of mJy~pc$^{-2}$) and \Ha\ (in units of
10$^{13}$~erg~cm$^{-2}$~s$^{-1}$~pc$^{-2}$) radial surface brightness distributions 
for discrete sources
at 8\mi\ (green diamonds), 24\mi\ (yellow triangles) and \Ha\ (black pluses).
{\it Bottom panel:} Fraction of the total fluxes due to emission from discrete sources
at 8\mi\ (green diamonds), at
24\mi\ (yellow triangles) and in the \Ha\ line (black pluses) as a function of radius.}
\label{fig:sources}
\end{figure}

Figure~\ref{fig:sources} illustrates the radial distributions of the
integrated flux in discrete sources, in bins of 0.48~kpc.
The upper panel shows that the IR flux at 8 and 24\mi\ contributed by
discrete sources declines radially more steeply than
the total (scale lengths of 0.86~$\pm$~0.06
and 0.88~$\pm$~0.13~kpc for 8 and 24\mi\ source number counts, respectively).
This is not the case for the \Ha\ sources, whose flux declines radially
with a scale length of 1.5~$\pm$~0.3~kpc, practically the same as the
total \Ha\ emission. Similar results are found for the contribution of UV sources.
Therefore, it is only the IR flux associated with discrete sources which shows a steep radial
decline. The very high IR diffuse fractions in the outer part of the SF disk are responsible for the
smooth decline of the light at IR wavelengths; no edge is seen, compared to the
\Ha\ and UV distributions which drop more sharply beyond 6~kpc.

The lower panel of Fig.~\ref{fig:sources}
displays the results for the fraction of emission contributed by discrete sources with
respect to the total. The \Ha\ fraction shows some
scatter but is nearly constant throughout the disk and equal to 0.5 on average;
this is in agreement with  \citet{2005ApJ...619L..67T} who,
however, found a slight increase towards the center.
The fraction of IR emission in discrete  sources is nearly
the same for the two wavelengths, but in the inner disc (up to 3.5~kpc)
nearly 40\% of the
emission is contributed by discrete sources,
while at larger radii they can account
for less than 20\%
\footnote{The discrepant point at 24\mi\ just beyond 5~kpc is due to the
contribution of IC\,133.}.
The drop of the IR flux associated with discrete sources is beyond 3 kpc, where
spiral arms appear to lose their strength. In the absence of strong perturbations, the
large dusty SF complexes disappear and smaller sources reside
in less opaque clumps (see also Corbelli et al. 2008, in preparation).

\subsection{Sources and diffuse TIR}

\begin{figure}
\includegraphics[width=\columnwidth]{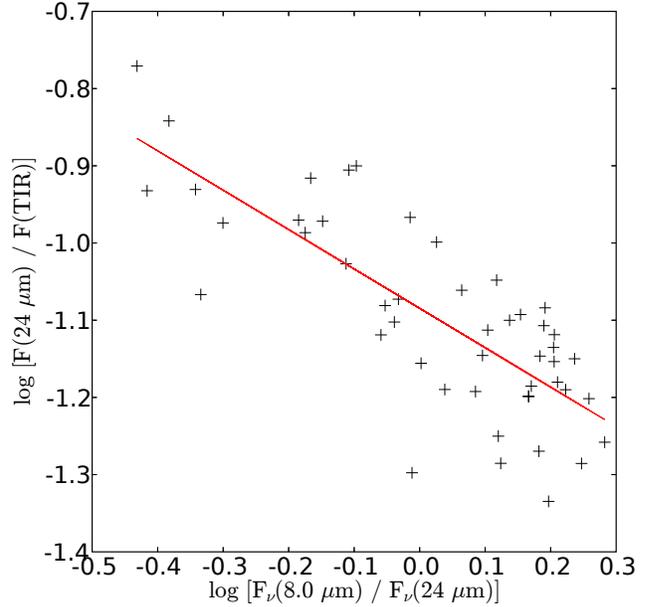}
\caption{The 24\mi\ to TIR luminosity ratio as a function of the 8\mi/24\mi\ flux ratio for 48
regions selected at 160\mi. The best linear fit to the data is shown by the plain (red) line.}
\label{fig:TIRsources}
\end{figure}

We will now derive the radial trend of the TIR
over the entire disk and of a sample of selected sources  on the 8 and 24\mi\
emissions. It is known that the relationship between 8, 24\mi, and TIR
luminosities depends on the properties of the galaxy considered and on the scale examined
\citep[e.g.][]{2005ApJ...628L..29E,2006ApJ...648..987P}.  
We will first examine if
the TIR emission in \mm, associated with selected sources or with the overall disk
emission, is correlated with the 8\mi\ emission.
For this purpose we extracted sources from the 160\mi\ image and selected only those
well fitted by closely circular PSF (ellipticity $<0.3$). Then we measured them,
within the aperture used at 160\mi, also at 8, 24, and 70\mi. We finally used the
formula in Eq.~\ref{eq:TIR} to estimate the TIR emission of 48 sources.
The 24\mi\ emission from discrete sources amounts to about 8\%
of their total TIR.
Figure~\ref{fig:TIRsources} shows the 24\mi\ to TIR luminosity ratio as a function of the
8 to 24\mi\ flux ratio. The best fit linear relation is :

\begin{equation} \label{eq:fit}
\log F({\rm TIR}) = \log F(24) + 1.08 + 0.51\ \log\left(\frac{F_\nu (8)}{F_\nu
(24)}\right) \quad .
\end{equation}

The $0.51\pm0.06$ coefficient is slightly lower than  what \citet{2005ApJ...633..871C}
derive in M\,51 using a fixed and larger aperture size but  
similar to what
\citet{2007ApJS..173..572T} derive for \mm\ using a sample of IR selected clumps
on scales of 160~pc. 
Figure~\ref{fig:TIRsources} shows that a correlation exists but also that the third term in
the above equation is small. This implies that the 8\mi\ emission from sources
depends on their TIR emission but does not contribute appreciably to it,
as expected
when dust is not very hot and PAH features dominate the 8\mi\ band.
Anyway, this result only holds for bright IR sources, those well
visible, resolved and isolated in the 160\mi\ map; 
the same might not apply to fainter sources.

Using the radial profiles we can also derive the fit to the same relation but for
the total emission (sources + diffuse component); we find :

\begin{equation} \label{eq:fitBin}
\log I_A({\rm TIR}) = \log I_A(24) + 1.13 + 0.06\ \log\left(\frac{I_{\nu A}  (8)}{I_{\nu A} (24)}\right) \quad .
\end{equation}

The correlation between the TIR and the 8\mi\ emissions is in this case nearly
absent, opposite to the trend found for selected IR sources.
The absence of such correlation 
in the diffuse emission, that is away from bright \hii\ complexes, 
is the result of a different radial trend of the 8\mi\ and of the IR emission at longer wavelength.

\begin{figure}
\includegraphics[width=\columnwidth]{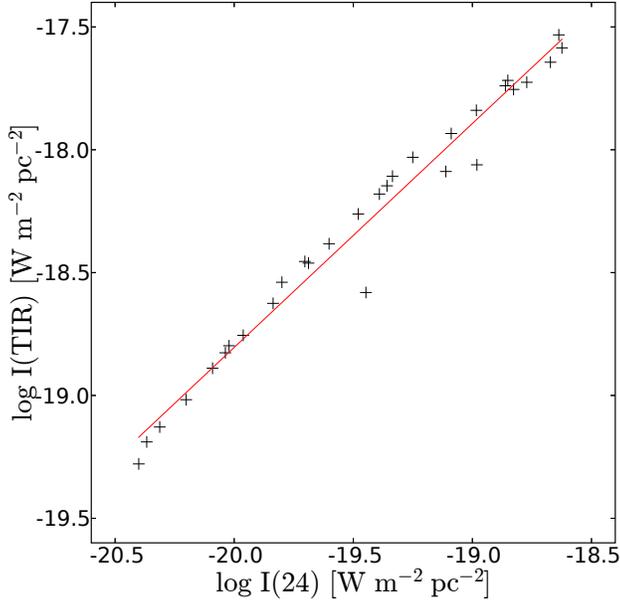}
\caption{TIR versus the 24\mi\ emission, from radial averages. The slope of the
best fitting linear relation between their logarithmic values (continuous line)
is 0.91.}
\label{fig:tir-24}
\end{figure}

Figure~\ref{fig:tir-24} shows the remarkably good correlation, with slope about
unity, between the TIR and the 24\mi\ fluxes when using elliptical averages.
This
does not necessarily imply that the powering source in the two bands is the same.
We shall see in the rest of this Section that
if two stellar populations of different ages are responsible for heating the
dust in the TIR and 24\mi\ band  respectively, these results
could imply that there has been no
variation in the radial dependence of the SFR between the two epochs.

\subsection{Origin of the unresolved 24\mi\ emission}

Discrete \hii\ regions in the innermost half of the SF disk
are clearly powering at least 40\% of dust radiation at 24 and 8\mi.
Then, what is left must be associated with dust spread out in
the disk and heated by the diffuse ISRF or it must be associated with
a ``sea'' of unresolved sources.
The fraction of dust associated with \hii\ regions decreases drastically
with radius not only because the SFR per unit area is declining radially
but also because they get dimmer in the IR beyond 3 kpc, as the drop in
the bottom panel of Figure 9 shows.
This raises the problem of the energy requirement at these wavelengths.
Grains associated with emission at 8 and
24\mi\ wavelengths are thought to be VSGs, not necessarily spatially
coincident with the large-size grains responsible for 
160\mi\ emission.
Additionally, at 8\mi\ the emission might have a large contribution from PAHs.
We therefore test two different hypotheses for
powering the diffuse emission at 24\mi:
{\it (a)} the interstellar radiation which is giving most of the
power by stochastic heating of VSGs;
{\it (b)} evolved AGB stars, mostly carbon-type stars which are
unresolved and have circumstellar dusty envelopes.

The first issue 
is the location of any unresolved 24\mi\ emission.
Radiation at this wavelength, in fact, must be produced {\it in situ}
and cannot be imputed to scattering of radiation emitted elsewhere.
This is easily verified by considering the extremely low scattering cross
section in the MIR.
For a standard MRN dust model
\citep{1977ApJ...217..425M}\footnote{$n(a)\propto a^{-3.5}$,
$a_{min}=0.001$\mi\ and $a_{max}=0.3$\mi,
silicate and carbonaceous grains in equal number.}
and the optical constants by \citet{1993ApJ...402..441L} for the silicate
material and by \citet{1996MNRAS.282.1321Z} for amorphous carbon,
we can estimate the scattering coefficient at 24\mi\ to be
about $1.4 \times 10^{-6}$ times the extinction coefficient
in the V band.
As we have seen, the face-on average $A_V$ value observed in the
disk is about 0.25, which translates in a total optical thickness of the disk,
after correcting  for scattering \citep{2005ApJ...619L..71B},
$\tau_V=0.35$. This means that, a 24\mi\ photon may travel several Mpc
in the \mm\ disk before being scattered into our line of sight by dust.
We note, incidentally, that the values observed for $A_V$ and for the total
(\hi +H$_2$) gas column density $N_{H}$
imply, for the dust model cited above, a mass density ratio
$\rho_{dust} / \rho_{H} = 3.9\times 10^{-3}$,
slightly lower than in our Galaxy where estimates span from $4\times 10^{-3}$
to $9\times 10^{-3}$ \citep{1992QB791.W45......}, and in overall agreement with
previous estimates, see \citet{1990A&A...236..237I} and references therein.

\subsubsection{Emission of Very Small Grains}

In the diffuse ISM, away from circumstellar environments,
stochastic, or single-photon, emission by VSGs is
thought to be the
main MIR continuum-emission mechanism \citep[e.g.][]{2007ApJ...657..810D}.
Precise computations are not easy to perform and in what follows
we illustrate a toy model aimed at a rough estimate of the VSG 
contribution to the diffuse 24\mi\ emission in \mm.

The emissivity of a single grain when
integrating over a time $\tau$ long enough to smooth out the
impulsive behaviour will be
\begin{equation}\label{eq:vsgemi}
\epsilon_\nu = \frac {\pi a^2\,Q^{abs}_\nu}{\tau}
\int_{0}^{\tau} B_\nu(T(t)) dt
\end{equation}
\noindent where $Q^{abs}_\nu$ is the absorption efficiency of the grain and
$B_\nu$ the Planck function at the grain temperature.
We will model the temperature behaviour with time $T(t)$
of a certain grain as a stochastic
sequence of identical, non-overlapping, flat-top events of height $T_g$
and duration $\Delta t$, and $T=0$ otherwise.
In each absorption event the grain will attain a temperature $T_g$
which depends on the energy of the incoming photon
and on the caloric properties of the grain. Given the photon,
independently of
the precise specific heat, the main factor will be the total number
of atoms in the lattice since $T_g \propto a^{-\delta}$,
where $\delta$ is the dimensionality.
Small grains are not expected to be spheres, they may
be flat or plain needles or, more probably,
a mixture of shapes and dimensions: we assume an average dimensionality
$\delta=2$, i.e., planar.
The grain will remain at temperature $T_g$ for the time $\Delta t$ it
takes to reradiate the absorbed energy.
For a given incoming photon energy,
$ \pi a^2\, \langle Q_{IR}^{abs} \rangle\, T_g^4\, \Delta t = const.$,
where $\langle Q_{IR}^{abs} \rangle$ is an average IR efficiency.
For the same illuminating field,
the number of events per grain during $\tau \gg \Delta t$ is
$N_{ev} \propto \tau \, \pi a^2 \, \langle Q_{UV}^{abs}  \rangle$,
where $\langle Q_{UV}^{abs} \rangle$ is an average efficiency
for the absorption of the heating photons.
For small grains of radius $a < 100$\,\AA,
we will always be in the Rayleigh regime and the absorption efficiency
$Q^{abs}_{\nu} \propto a$.
The characteristic grain temperatures we are interested in
are those for which
$\lambda F_\lambda$ peaks at 24\mi, that is $T_g \simeq 150$\,K.
Around this temperature, the Planck function
at 24\mi\ can be approximated, as a
power law of $T$ by $B_{24}(T)\propto T^{3.9}$.
Putting all this together into Eq.\,\ref{eq:vsgemi},
it is easily derived that the 24\mi\ emissivity of the
single grain varies as $\epsilon_{24} \propto a^{3.2}$.
This is in remarkably good agreement with simulations
of the VSG emission in the range 10 to 50\,\AA\ of radius
\citep{2003pid..book.....K}. Obviously this relation will hold only
in the range of grain radii which
can be heated up to $\sim 150$\,K by single-photon absorptions.
The grains to be taken in consideration when heated
by photons in the range $900 < \lambda < 5000$\,\AA\
are those containing roughly between 100 and 600 atoms,
which translates into $15 < a < 40$\,\AA.

Following this line, upon averaging in time and composition,
we estimate the emissivity of
a $a=40$\,\AA\ grain in the local ISRF to be
$\epsilon_{24} = 4.3\times 10^{-31}$
 \,erg s$^{-1}$Hz$^{-1}$sr$^{-1}$.
If we normalize the grain
emission to this value and then integrate on the MRN size distribution
between 10 and 50\,\AA\ we can derive a volume emissivity and,
given the observed gas-to-dust ratio, obtain the surface
brightness at 24\mi\ contributed by VSGs which,
in this scheme, will be simply
proportional to the gas column density and to the ISRF:
$\sigma_{24}^{VSG}=6.0\times 10^{-6} \sigma_H U $\,mJy pc$^{-2}$,
where $ \sigma_H$ is the H-gas column density in \msun\,pc$^{-2}$,
and $U$ the ratio between the actual ISRF and the one in the
solar neighborhood\footnote{In the same units \citet{2007ApJ...657..810D}
obtain
$\sigma_{24}^{VSG} \simeq 2.0\times 10^{-5} \sigma_H U $\,mJy pc$^{-2}$.}.

In order to estimate the factor $U$,
we derive in Appendix\,\ref{app} an approximate solution
for a disk-like atmosphere which allows us to derive the field at
the midplane, given the surface brightness seen by an external
observer. We will use this method to estimate the \mm\ ISRF at various wavelengths
and compare it to the local values in the Galaxy \citep{2001ApJS..134..263W}.

Since the 3.6\mi\ waveband is thought to directly measure
the evolved stars photospheric emission, we may use it
as a proxy of the overall stellar content and
therefore of the ISRF itself. Given the stellar population responsible
for the 3.6\mi\ emission, we will postulate a ratio between dust
and stars scale heights $\beta=1/3$; in \mm\ at this wavelength
we estimate the optical half-thickness $\tau_0/\mu\simeq 0.02$.
With these assumptions,
at 4\,kpc in the outer disk of \mm, the observed 3.6\mi\
surface brightness implies a midplane ISRF which is
0.31 times the local Galactic value.
Alternatively, since the VSG stochastic emission is powered mainly by
UV photons, we may directly estimate the ISRF from the
\galex\ data. We assume $\beta=1/1$ and $\tau_0/\mu\simeq 0.9$ for both
the NUV and FUV emission and in both cases we consistently derive,
at 4\,kpc, an
ISRF 0.50 times the one in the solar neighborhood,
in overall agreement with the 3.6\mi\ result.
If we then assume that the
ISRF in the outer \mm\ disk is about half the local Galactic value,
we find that the predicted VSGs emission falls short
by a factor $>10$, of the diffuse 24\mi\ flux actually measured\footnote{or
a factor $>3$ when adopting the more generous emissivities by
\citet{2007ApJ...657..810D}.}.
Although exemplified here at a fiducial radius of 4\,kpc,
this happens to be true at all radii,
as well as for the total, unresolved 24\mi\ luminosity.
We have verified that this result depends weakly on the
limits used for the range of grain radii and on the slope of
grain size distribution. Also, the VSG emission of
spherical grains will be lower than what we have here estimated
for flat ($\delta = 2$) geometries.

\subsubsection{Carbon stars}

To better understand the origin of the diffuse 24\mi\ emission,
we focus on a class of intermediate-age stars, the carbon stars. 
Carbon-rich giants, or carbon stars (CSs) are stars of low
to intermediate initial mass, 0.8 to 4\msun\ with extended
dusty circumstellar envelopes mainly found in their asymptotic
giant branch (AGB) evolutionary phase \citep{2005ARA&A..43..435H}.
460 carbon
stars have been observed using the {\it Spitzer} infrared
spectrograph (IRS) in the Magellanic Clouds by \citet{2007MNRAS.376..313G}
and half of them in the LMC, an environment not
dissimilar from the \mm\ disk.
While their $\lambda F_{\lambda}$
peaks around 8\mi, they still show rather strong emission at
longer wavelengths, with a $\lambda F_{\lambda}$ at 100\mi\
comparable to the one in the J bandpass. The observed spread in
luminosity covers about 1.7~dex with the most luminous ones
being the reddest, as expected in the case of AGB stars;
when set
at the distance of \mm, the average 24\mi\ flux of a single CS
is $F_{24}=0.21$~mJy.

In order to estimate the possible
contribution to the MIR emission by CSs we will make use of the
Padova evolutionary tracks by \citet{2000A&AS..141..371G} for a metallic
content Z=0.004\footnote{The tracks do not cover the full AGB
phase but only up to the onset of thermal pulses. Under this
respect, our estimates constitute a lower limit to the CS
contribution; due to the short duration of the pulsing phase,
the correction is of little importance.}.
It is during the shell
helium-burning phase that stars climb the red supergiant branch
in the HR diagram by increasing their luminosity and lowering
the effective temperature.
The duration of this AGB phase
varies with the initial mass, from about $10^7$ to $10^6$\,yr
in the CS mass range.
Similarly, the luminosity attained on
the AGB depends on mass but three quantities are roughly
constant with mass:
(1) the effective temperature range from
$T_{ini} \simeq 4500$ to $T_{fin} \simeq 3800$\,K;
(2) the ratio between final and initial luminosity
$L_{fin}/L_{ini}\simeq 13$ (that is, in log-log,
different masses evolve along identical, vertically displaced,
parallel tracks);
and (3) the bolometric energy
release $\int_{AGB}L\,dt \simeq 4.5\times 10^9$\lsun\,yr.

If $L(m,t_A)$ is the bolometric luminosity of a CS of initial mass $m$ after
a time $t_A$ since the entrance in AGB phase, we can define
a time-averaged luminosity at mass $m$:

\begin{equation} \label{eq:timeavlum}
\langle L(m) \rangle = \frac{\int_{\,
0}^{\,\tau_A(m)} L\,(m,t_A)\,dt_A}{\tau_A(m)}
\end{equation}

\noindent where $\tau_a(m)$ is the AGB phase duration for mass $m$.
Given an Initial Mass Function, we can derive an overall mean luminosity
for all CSs:

\begin{equation} \label{eq:totavlum}
\langle \langle L \rangle \rangle =
\frac
{\int_{\,0.8\,{\rm M}_\odot}^{\,4.0\,{\rm M}_\odot} \langle L(m) \rangle \, m^{-\alpha} \, dm}
{\int_{\,0.8\,{\rm M}_\odot}^{\,4.0\,{\rm M}_\odot} m^{-\alpha} \, dm}
\end{equation}

\noindent where $\alpha$ is the exponent of the IMF power law. Hereafter
we adopt $\alpha=2.3$ for $0.5\le m \le 100$\,M$_\odot$ \citep{1955ApJ...121..161S} and 1.3 for
$0.1\le m \le 0.5$\,M$_\odot$ \citep{2001MNRAS.322..231K}.
For this IMF and the Padova tracks, we obtain $\langle \langle L \rangle \rangle = 580$\lsun.

As we said, based on the observations in LMC, we know the average expected 24\mi\ flux from
a CS in \mm, and we may think of this as a mass and time average over the whole CS population,
as outlined above for the luminosity, that is
$\langle \langle F_{24} \rangle \rangle = 0.21$\,mJy.
We also noted that the effective temperature range of CS is rather restricted and quite
similar for all the relevant masses.
Given this circumstance, we may assume
that with a good approximation
$\langle \langle F_{24} \rangle \rangle \propto \langle \langle L \rangle \rangle$
with a proportionality constant
$\langle\langle K_{24-L}\rangle\rangle =\langle\langle F_{24}\rangle\rangle / 
\langle\langle L \rangle\rangle = 3.65 \times 10^{-4}$\,mJy\lsun$^{-1}$.

At this point, if $F_{24}(m,t_A)$ is the flux from the single CS, from a given
region we expect a flux

\begin{equation}\label{eq:sigmacs}
\Sigma_{24} = \int_ {\,0.8\,{\rm M}_\odot}^{\,4.0\,{\rm M}_\odot}
\dot{N}(m) \, \left\{\int_{\,0}^{\,\tau_A(m)} F_{24}(m,t_A) \,dt_A \right\} \,dm
\end{equation}

\noindent where $\dot{N}(m)$ is the differential star-formation rate in that region,
admitting it is constant with time (both integrally and in mass distribution).
If we now assume that $F_{24}(m,t_A) = \langle\langle K_{24-L}\rangle\rangle L\,(m,t_A)$,
the theoretical tracks can be used to evaluate Eq.~\ref{eq:sigmacs} and we obtain
$\Sigma_{24} = 2.05 \times 10^5 \dot{M}_T$\,mJy, where $\dot{M}_T$ is the total
(from 0.08 to 100\msun) SFR in the region of interest
in \msun\ yr$^{-1}$. Note that, despite the rather strong assumption of proportionality
between $F_{24}(m,t_A)$ and $L\,(m,t_A)$, the final result relies on estimates
of $\int L\,dt$ along the AGB phase which, as said above, are quite robust
and constant.

It turns out that the SFR required to explain the 24\mi\ unresolved emission in \mm\
with the sole contribution of CSs
is quite compatible with what is computed by the usual SFR indicators,
as will be seen in the following.
Note also that the most luminous CSs
could well have been counted among the faint discrete sources, and that a small
fraction of carbon stars might not have dusty shells
\citep{2007ApJ...664..850M}. As a result the 24\mi\ unresolved emission due to CS
should give a lower SFR than the effective one.
The total unresolved 24\mi\ flux of \mm, $\sim 35$\,Jy, 
translates into a
global SFR of  $0.17$\msun\ yr$^{-1}$;
the surface brightness in the outer disk at 4\,kpc is
$\sim 4 \times 10^{-4}$~mJy~pc$^{-2}$
requiring $\dot{M}_T \simeq 2.0$\msun\ Gyr$^{-1}$\,pc$^{-2}$.

\section{Star formation rates} \label{sec:sfr}

How well tracers at different
wavelengths measure the actual SFR has been examined
by a number of papers based on \spitzer\ data
\citep[e.g.][for M\,51]{2005ApJ...633..871C, 2007ApJ...671..333K}.
In this Section, we compare the SFRs derived from emission at different wavelengths, as a
function of the galactocentric radius (in bins of 0.24~kpc) and integrated over the whole SF disk.
To compute the SFR from \Ha\ and UV luminosities corrected for extinction (see
Sect.~\ref{sec:radial})
we use the stellar population synthesis model in Starburst~99 \citep{1999ApJS..123....3L},
for a continuous star-formation model with solar metallicity and the IMF
cited in the previous Section.

\subsection{SFR from \Ha\ and UV wavelengths}

The conversion between \Ha\ luminosities and SFR given by the stellar population model
is

\begin{equation} \label{eq:SFRHa}
{\rm SFR(H}\alpha)\ [M_\odot\ {\rm yr}^{-1}] = 8.3 \times 10^{-42}\ L({\rm H}\alpha)\ [{\rm
erg~s}^{-1}] \quad .
\end{equation}

This is within 5\% of the value given by \citet{1998ARA&A..36..189K},
showing that the SFR dependence on \Ha\ is rather insensitive
to the assumptions about the lower-mass form of the IMF.
The radial profile is shown in Fig.~\ref{fig:sfr}
and the total integrated SFR over the whole SF disk
($R<7$~kpc) is SFR(\Ha)~=~0.35~$M_\odot$~yr$^{-1}$ which
correspond to the average SFR over about the last 10~Myr.
To trace the SFR over a wider timescale, namely in the last 100~Myr, we convert the FUV
luminosity according to the same population synthesis model at  1550~\,\AA\ as:

\begin{equation} \label{eq:SFRFUV}
{\rm SFR(FUV)}\ [M_\odot\ {\rm yr}^{-1}] = 8.8 \times 10^{-44} L(\rm {FUV)\ [erg~s}^{-1}] \quad .
\end{equation}

The above conversion (equivalent to
$1.7 \times 10^{-28} L_\nu(\rm {FUV)\ [erg~s}^{-1}~{\rm Hz}^{-1}]$)
is also similar to the one given by \citet{1998ARA&A..36..189K}
(although we are using a slightly different IMF and wavelength range).
The radial profile of the SFR derived from the FUV emission is shown
in Fig.~\ref{fig:sfr} and the SFR integrated over the SF disk is
SFR(FUV)~=~0.55~$M_\odot$~yr$^{-1}$,
higher that the one derived from \Ha.
The SFR declines radially with a scale length of 2.2~$\pm$~0.1~kpc in the FUV and
1.8~$\pm$~0.1~kpc in \Ha,
a marginal difference considering the uncertainties.
However there are two reasons why the two might indeed be different:
one is because we are sampling two different SF epochs; the second is that \Ha\
depends on the interstellar medium volume density 
and a radial decrease of this might suppress
\Ha\, and also induce a sharper \Ha\ SF edge \citep{2006ApJ...636..712E}.

\subsection{SFR from IR wavelengths}

While in dusty environments IR-based SF tracers are quite reliable,
in \mm\ this assumption is not obvious and requires caution in interpretation.
Although within the single SF region most of the dust
emission is powered by young stars, the situation is more complex
when we integrate over the whole disk which, as said before,
has an important diffuse fraction.
In \mm\ the emission at 70 and 160\mi\ can be
powered by the UV in the diffuse ISRF or by single \hii\ regions:
both cases are related to the presence of stars of young to moderate age.
We expect, therefore, the FIR luminosity to be
a valid SF tracer and yield a SFR similar to the one based on FUV data,
once corrected for extinction.
If we compute the FIR luminosity by fitting an optically thin
$(\kappa_\lambda \propto \lambda^{-2})$ thermal
spectrum to the 70 and 160\mi\ elliptical averages
and then integrate this between 40 and
1100\mi\ (see Sect.~\ref{sec:gasDust}), we find that it correlates well
with the TIR luminosity computed via Eq.~\ref{eq:TIR}.
The average ratio between
the TIR and the FIR luminosities is $1.40 \pm 0.04$ across the whole disk.
The FIR emission thus modeled, integrated over the disk ($< 7$~kpc),
amounts to $4.6 \times 10^{42}$~erg~s$^{-1}$.

If we choose the extinction corrected FUV as a reference for the SFR,
we can derive the conversion factors for the other wavebands.
We did this by using a $\chi^2$ minimization over the set of radial bins
within 6~kpc, to avoid the most noisy data.
For the FIR we obtain:

\begin{equation} \label{eq:SFRFIR2}
{\rm SFR(FIR)}\ [M_\odot~{\rm yr}^{-1}] = 13 \times 10^{-44} L({\rm FIR)\ [erg~s}^{-1}] \quad ,
\end{equation}

\noindent which is in the upper range of the values found by \citet{1996A&A...306...61B} for late-type
galaxies based on IRAS and UV flux measurements.
A relatively high conversion factor is indicative of a relatively low dust content.

Using the same technique but for the TIR fluxes, we find that
in order to match the FUV SFR:

\begin{equation} \label{eq:SFRTIR}
{\rm SFR(TIR)}\ [M_\odot~{\rm yr}^{-1}] = 8.8 \times 10^{-44} L({\rm TIR)\ [erg~s}^{-1}] \quad .
\end{equation}

\begin{figure}
\includegraphics[width=\columnwidth]{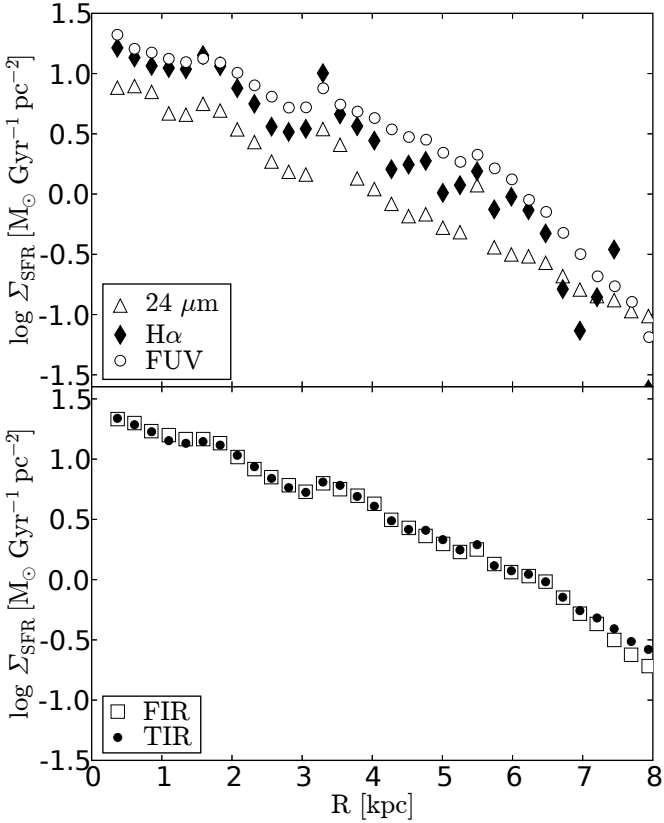}
\caption{SFR per square parsec as a function of radius using Eq.~\ref{eq:SFRFUV} for converting into
SFR the
FUV luminosity (open circles), Eq.~\ref{eq:SFR24} for the 24\mi\ luminosity
(open triangles)
Eq.~\ref{eq:SFRHa} for the \Ha\ luminosity (filled diamonds) ({\it upper panel}),
Eq.~11 for the FIR
luminosity (open squares) and Eq.~\ref{eq:SFRTIR} for the TIR luminosity (filled circles)
({\it bottom panel}).}
\label{fig:sfr}
\end{figure}

As seen in Fig. \ref{fig:sfr}, the FIR- and TIR-based SFR radial profiles show a
smoother drop at the SF disk edge possibly due to their origin from more 
diffuse components, dust, and general ISRF.

If one is convinced that also the 24\mi\ is powered by young stars,
then normalizing the SFR
based on the 24\mi\ emission to the FUV SFR, one obtains:

\begin{equation} \label{eq:SFR242}
{\rm SFR(24)}\ [M_\odot~{\rm yr}^{-1}] = 1.2 \times 10^{-42} L({\rm 24)\ [erg~s}^{-1}] \quad .
\end{equation}

This factor is about three times higher with respect to
the relation given by \citet{2005ApJ...632L..79W}
\citep[see also][]{2001ApJ...554..803Y}, which was derived by comparing
IR and radio emission in the \spitzer\ First Look Survey Galaxies.
They find, in fact:

\begin{equation} \label{eq:SFR24}
{\rm SFR(24)}\ [M_\odot~{\rm yr}^{-1}] = 3.9 \times 10^{-43} L({\rm 24)\ [erg~s}^{-1}] \quad .
\end{equation}

Using this latter relation the global SFR from 24\mi\ emission in \mm\ is only 0.2~$M_\odot$~yr$^{-1}$
and the relative SFR per unit area is shown in Fig.~\ref{fig:sfr}.
Eq.~\ref{eq:SFR24} clearly underestimate the SFR. In the inner regions
the 24\mi\ and \Ha\ SFRs follow the pattern of the galaxy: they are enhanced along
the spiral arms and decrease monotonically away from the center. 
Beyond 4~kpc the 24\mi, emission fraction linked to
discrete sources drops. Diffuse emission dominates which,
as shown in the previous Section, can  be powered by more evolved
stars and linked to the SFR about 1~Gyr ago rather than to recent episodes of SF.
This explains the absence of any drop at the SF disk edge due to stellar diffusion
of the more evolved population.
Hence, the 24\mi\ emission in late-type galaxies like \mm, with a few very large
SF complexes and a large diffuse component,
depends upon the presence of dusty envelopes around
evolved stars. If the contribution in hot discrete sources is
removed, it can give a lower limit to the SFR about 1~Gyr ago: the exact rate
depends upon the fraction of evolved stars which have developed a dusty envelope.

\section{Summary and conclusions} \label{sec:conclusion}

This second paper of the series focuses on various SFR diagnostics across the disk
of the local spiral galaxy \mm, and on the origin of the large IR diffuse emission.
We investigate the radial trends of the IR emission at various wavelengths
using \spitzer\ data, complementing these with \galex\ data for the UV,
with \Ha\ data, and with data on the atomic and molecular gas distributions.
Hereafter, we summarise our main results.

\begin{itemize}
\item
The \spitzer\ radial profiles at 3.6, 4.5, 24, and 70\mi\ are
consistent with a single exponential
scale length throughout the star-forming disk. This is $\sim$1.55~kpc for
the (old) stellar population emission and
$\sim$1.75~kpc for the dust emission at 24 and 70\mi.
Two different slopes for the inner and outer star-forming disk
are found instead at 5.8, 8.0 and 160\mi.
At these wavelengths the IR brightness
declines more steeply beyond 3.5~kpc and drops at the edge of the star-forming disk.
The scale lengths more directly linked to SF diagnostics (\Ha, FUV and NUV) are 
about 2~kpc, comparable to the IR ones in the inner disk, but longer than these 
in the outer star-forming disk.
This implies that the dust column density is decreasing beyond 3.5 kpc, 
a trend confirmed by our estimate of the extinction from the
TIR to FUV ratio. The 8\mi\ emission is dominated by PAH features
which show instead a faster decline with radius.

\item
Within the star-forming disk, that is within 7~kpc of radius,
we estimate the TIR, FUV, NUV, and \Ha\ luminosities to be
$6.8 \times 10^{42}$, $7.0 \times 10^{42}$, $6.3 \times 10^{42}$, 
and $4.2 \times 10^{40}$~erg~s$^{-1}$ respectively
after correcting for extinction. Extinction corrections are small and
reflect the rather low dust content of the galaxy ($A_{\rm V} \sim 0.25$, 
for stellar continuum averaged in rings of width 0.24~kpc).
Extinction can be
enhanced locally (cores of \hii\ regions, for instance);
however, the absence of large dusty complexes beyond 3~kpc, likely due to the
weakness of the spiral pattern, implies that at large radii SF proceeds in smaller
clumps with low extinction.
Dust opacity derived from the 160\mi\ emission under the assumption of single
temperature component has a steeper radial decline than dust opacity inferred
via the extinction model. 

\item
A substantial amount of the emission of the \mm\ galaxy is in diffuse form at
several wavelengths. 
The FUV diffuse fraction is equal to about 0.65 and does not depend on radius. 
The \Ha\ diffuse fraction is more scattered but nearly constant throughout
the disk of the galaxy and equals to 0.5 on average.
In the IR, at 8 and 24\mi, about 40\% of the total emission of the inner disk comes from
discrete sources while this contribution drops to 20\% in the outer disk.
The dust temperature, estimated from the 70 and 160\mi\ emission, decreases 
from 25 to 21~K from the center to 4~kpc and,
after a plateau, it might increase at the edge of the star-forming disk.
The average ratio between the TIR and FIR luminosities is $1.40\pm 0.04$ 
across the whole disk.

\item 
We examine the SFR of \mm\ from various tracers
(extinction corrected \Ha\ and FUV, TIR, FIR, 24\mi). We estimate
the total integrated SFR to be $0.45\pm 0.1$~$M_\odot$~yr$^{-1}$.
From the \Ha\ luminosity (which accounts for SF in the last $< 10$~Myr) we
infer a lower value, SFR(\Ha)~=~0.35~$M_\odot$~yr$^{-1}$, while the FUV emission
(which accounts for SF in the last $< 100$~Myr)
gives SFR(FUV)~=~0.55~$M_\odot$~yr$^{-1}$;
the difference is small with respect to the uncertainties, 
so that we cannot conclude definitively that SFR is declining in the last 100~Myrs.
If we calibrate the SFR from FIR and TIR emission through the values from UV
data, we obtain conversion factors in the upper range
of what is generally found for late-type galaxies,
likely due to the low dust content of \mm.

\item
A SFR diagnostic based on the 24\mi\ emission in \mm\ gives a conversion
factor more than three times higher than the standard ones.
To estimate the validity of the 24\mi\ emission as a SFR tracer away from \hii\ complexes,
we investigated the nature of its high diffuse fraction.
We find that VSGs emission,
as powered by the diffuse ISRF, accounts for only 10\% of the diffuse 24\mi\ emission,
while unresolved, AGB stars of intermediate mass, Carbon Stars, through their
dusty circumstellar envelopes, can provide most of the diffuse 24\mi\ emission.
If indeed evolved stars contribute substantially to the 24\mi\ brightness,
emission at 24\mi\ can trace SFR about 1~Gyr ago.
In this paper we only give a lower limit to it
of about 0.20~$M_\odot$~yr$^{-1}$. This estimate takes into account only the diffuse
24\mi\ emission and does not consider the
most luminous CSs, which might be resolved as discrete sources, as well as the fraction of
CSs without dusty shells. With the help of future surveys of evolved stars in \mm, it
should soon be possible to complement our study and obtain a better view of the 
time evolution of the SFR in the Triangulum galaxy.

\end{itemize}

\appendix
\section{A simplified exponential atmosphere} \label{app}
We model the disk as a plane parallel medium symmetric respect to midplane.
In case of isotropic extinction and emissivity,
at vertical distance $z$ from midplane, the transfer of the
intensity $I_\nu\,(z,\theta)$ along a direction
at angle $\theta$ from the vertical will follow
\begin{equation}\label{eq:transfer1}
\mu \,\,\frac{dI_\nu \,(z,\theta)}{\chi_\nu\,(z) \, dz} =
S_\nu \,(z) - I_\nu \,(z,\theta) \quad ,
\end{equation}
as usual $\mu=cos\,\theta$, $\chi_\nu$ is the volume extinction coefficient, and
$S_\nu$ is the source function;
$S_\nu\,(z)=\epsilon_\nu\,(z)/\chi_\nu\,(z)$,
where $\epsilon_\nu\,(z)$ is the volume emissivity.
If $\tau_\nu$ is the {\it vertical} optical thickness
$\tau_\nu\,(z)=\int_z^\infty\chi_\nu\,(z')\,dz'$ , then
\begin{equation}\label{eq:transfer2}
\mu \,\,\frac{dI_\nu \,(\tau,\mu)}{d\tau} =
I_\nu \,(\tau,\mu) - S_\nu \,(\tau) \quad .
\end{equation}
If we assume, though not strictly required,
that the intensity at midplane is also isotropic, $I_\nu\,(z=0,\theta)=J_\nu(0)$,
then the intensity emerging from the atmosphere
($z=+\infty, \, \tau=0, \, \mu> 0$) will be
\begin{equation}\label{eq:outgoing}
I_\nu(0,\mu)=J_\nu(0) \, e^{-\tau_0/\mu} +
\int_0^{\tau_0} S(\tau) \, e^{-\tau/\mu} \, \frac{d\tau}{\mu} \quad  ,
\end{equation}
where $\tau_0=\tau(z=0)$ is the optical half-thickness of the disk. At the same time,
at midplane ($z=0, \,\tau=\tau_0, \,\mu< 0$), if there is no incoming radiation from
outside the disk, we have
\begin{equation}\label{eq:incoming}
I_\nu(\tau_0,\mu)=J_\nu(0)=
\int_0^{\tau_0} S(\tau) \, e^{-(\tau_0-\tau)/\mu} \, \frac{d\tau}{\mu} \quad ,
\end{equation}
If the extinction and emissivity volume coefficients depend
exponentially on $z$, so will be $S_\nu(z)$ and the integrals in
Eq.s\,\ref{eq:outgoing},\,\ref{eq:incoming} can be evaluated in terms
of Incomplete Gamma functions $\gamma\,(\,\beta,\, \pm \, \tau_0/\mu)$,
where $\beta$ is the ratio between the extinction and emissivity scale heights,
$\beta = h_{ext}/h_{em}$.
By comparing the two solutions we find the emergent intensity $I_\nu(\tau=0,\mu)$
in terms of the one at midplane $J_\nu(0)$, and vice versa:
\begin{equation}\label{eq:solution}
I_\nu(0,\mu)=J_\nu(0) \, e^{-\tau_0/\mu} \, + \, J_\nu(0) \,
\frac{e^{\tau_0/\mu} \, \gamma \,(\, \beta,\, \tau_0/\mu)}
{(-1)^\beta \, \gamma \, (\, \beta,\, - \, \tau_0/\mu)}
\quad .
\end{equation}
Some numerical values of $J_\nu(0)\,/I(0,\mu)$,
the ratio between midplane and emergent intensity are
shown in Table\,\ref{tab:expatm}. For $\beta=1$ we recover the
values for the particular case of constant $S_\nu$.
\begin{table}
\begin{center}
\caption{Midplane to emergent intensity ratio $J_\nu(0)\,/I(0,\mu).$} \label{tab:expatm}
\begin{tabular}{c c c c}
\hline \hline
 \hfill\vline & & & \\
 $\tau_0/\mu$ \hfill\vline & $\beta=\frac{1}{1}$ & $\beta=\frac{1}{2}$ & $\beta=\frac{1}{3}$ \\
 \hfill\vline & & & \\\hline

0.1 \hfill\vline & 0.53  & 0.52 & 0.51 \\
0.6 \hfill\vline & 0.65  & 0.56 & 0.53 \\
1.6 \hfill\vline & 0.83  & 0.54 & 0.42 \\
3.6 \hfill\vline & 0.97  & 0.36 & 0.22 \\
6.6 \hfill\vline & 1.00  & 0.24 & 0.12 \\

\hline
\end{tabular}
\end{center}
\end{table}

\begin{acknowledgements}
We would like to thank Rene Walterbos for providing us the \Ha\ image of \mm,
David Thilker for the GALEX-UV profiles and George Helou for discussions and advice.
We also thank the anonymous referee whose insightful comments improved the clarity of the paper.
The work of
S.~V. is supported by a INAF--Osservatorio Astrofisico di Arcetri fellowship. The \spitzer\ Space
Telescope is
operated by the Jet Propulsion Laboratory, California Institute of Technology, under contract with
the
National Aeronautics and Space Administration. This research has made use of the NASA / IPAC
Extragalactic
Database, which is operated by JPL /Caltech, under contract with NASA.
\end{acknowledgements}

\bibliography{astroph}

\begin{thebibliography}{75}
\expandafter\ifx\csname natexlab\endcsname\relax\def\natexlab#1{#1}\fi

\bibitem[{{Bendo} {et~al.}(2008){Bendo}, {Draine}, {Engelbracht}, {Helou},
  {Thornley}, {Bot}, {Buckalew}, {Calzetti}, {Dale}, {Hollenbach}, {Li}, \&
  {Moustakas}}]{2008arXiv0806.2758B}
{Bendo}, G.~J., {Draine}, B.~T., {Engelbracht}, C.~W., {et~al.} 2008, ArXiv
  e-prints, 806

\bibitem[{{Bertin} \& {Arnouts}(1996)}]{1996A&AS..117..393B}
{Bertin}, E. \& {Arnouts}, S. 1996, \aaps, 117, 393

\bibitem[{{Bianchi} {et~al.}(2005){Bianchi}, {Thilker}, {Burgarella},
  {Friedman}, {Hoopes}, {Boissier}, {Gil de Paz}, {Barlow}, {Byun}, {Donas},
  {Forster}, {Heckman}, {Jelinsky}, {Lee}, {Madore}, {Malina}, {Martin},
  {Milliard}, {Morrissey}, {Neff}, {Rich}, {Schiminovich}, {Siegmund}, {Small},
  {Szalay}, {Welsh}, \& {Wyder}}]{2005ApJ...619L..71B}
{Bianchi}, L., {Thilker}, D.~A., {Burgarella}, D., {et~al.} 2005, \apjl, 619,
  L71

\bibitem[{{Buat} \& {Xu}(1996)}]{1996A&A...306...61B}
{Buat}, V. \& {Xu}, C. 1996, \aap, 306, 61

\bibitem[{{Calzetti}(2001)}]{2001PASP..113.1449C}
{Calzetti}, D. 2001, \pasp, 113, 1449

\bibitem[{{Calzetti} {et~al.}(2005){Calzetti}, {Kennicutt}, {Bianchi},
  {Thilker}, {Dale}, {Engelbracht}, {Leitherer}, {Meyer}, {Sosey}, {Mutchler},
  {Regan}, {Thornley}, {Armus}, {Bendo}, {Boissier}, {Boselli}, {Draine},
  {Gordon}, {Helou}, {Hollenbach}, {Kewley}, {Madore}, {Martin}, {Murphy},
  {Rieke}, {Rieke}, {Roussel}, {Sheth}, {Smith}, {Walter}, {White}, {Yi},
  {Scoville}, {Polletta}, \& {Lindler}}]{2005ApJ...633..871C}
{Calzetti}, D., {Kennicutt}, Jr., R.~C., {Bianchi}, L., {et~al.} 2005, \apj,
  633, 871

\bibitem[{{Cannon} {et~al.}(2006){Cannon}, {Walter}, {Armus}, {Bendo},
  {Calzetti}, {Draine}, {Engelbracht}, {Helou}, {Kennicutt}, {Leitherer},
  {Roussel}, {Bot}, {Buckalew}, {Dale}, {de Blok}, {Gordon}, {Hollenbach},
  {Jarrett}, {Meyer}, {Murphy}, {Sheth}, \& {Thornley}}]{2006ApJ...652.1170C}
{Cannon}, J.~M., {Walter}, F., {Armus}, L., {et~al.} 2006, \apj, 652, 1170

\bibitem[{{Cioni} {et~al.}(2008){Cioni}, {Irwin}, {Ferguson}, {McConnachie},
  {Conn}, {Huxor}, {Ibata}, {Lewis}, \& {Tanvir}}]{2008arXiv0805.1143C}
{Cioni}, M.~.~L., {Irwin}, M., {Ferguson}, A.~M.~N., {et~al.} 2008, ArXiv
  e-prints, 805

\bibitem[{{Corbelli}(2003)}]{2003MNRAS.342..199C}
{Corbelli}, E. 2003, \mnras, 342, 199

\bibitem[{{Corbelli} \& {Salucci}(2000)}]{2000MNRAS.311..441C}
{Corbelli}, E. \& {Salucci}, P. 2000, \mnras, 311, 441

\bibitem[{{Corbelli} \& {Schneider}(1997)}]{1997ApJ...479..244C}
{Corbelli}, E. \& {Schneider}, S.~E. 1997, \apj, 479, 244

\bibitem[{{Corbelli} \& {Walterbos}(2007)}]{2007ApJ...669..315C}
{Corbelli}, E. \& {Walterbos}, R.~A.~M. 2007, \apj, 669, 315

\bibitem[{{Dale} {et~al.}(2005){Dale}, {Bendo}, {Engelbracht}, {Gordon},
  {Regan}, {Armus}, {Cannon}, {Calzetti}, {Draine}, {Helou}, {Joseph},
  {Kennicutt}, {Li}, {Murphy}, {Roussel}, {Walter}, {Hanson}, {Hollenbach},
  {Jarrett}, {Kewley}, {Lamanna}, {Leitherer}, {Meyer}, {Rieke}, {Rieke},
  {Sheth}, {Smith}, \& {Thornley}}]{2005ApJ...633..857D}
{Dale}, D.~A., {Bendo}, G.~J., {Engelbracht}, C.~W., {et~al.} 2005, \apj, 633,
  857

\bibitem[{{Dale} \& {Helou}(2002)}]{2002ApJ...576..159D}
{Dale}, D.~A. \& {Helou}, G. 2002, \apj, 576, 159

\bibitem[{{Deul} \& {van der Hulst}(1987)}]{1987A&AS...67..509D}
{Deul}, E.~R. \& {van der Hulst}, J.~M. 1987, \aaps, 67, 509

\bibitem[{{Devereux} {et~al.}(1997){Devereux}, {Duric}, \&
  {Scowen}}]{1997AJ....113..236D}
{Devereux}, N., {Duric}, N., \& {Scowen}, P.~A. 1997, \aj, 113, 236

\bibitem[{{Draine} \& {Lee}(1984)}]{1984ApJ...285...89D}
{Draine}, B.~T. \& {Lee}, H.~M. 1984, \apj, 285, 89

\bibitem[{{Draine} \& {Li}(2007)}]{2007ApJ...657..810D}
{Draine}, B.~T. \& {Li}, A. 2007, \apj, 657, 810

\bibitem[{{Elmegreen} \& {Hunter}(2006)}]{2006ApJ...636..712E}
{Elmegreen}, B.~G. \& {Hunter}, D.~A. 2006, \apj, 636, 712

\bibitem[{{Engargiola} {et~al.}(2003){Engargiola}, {Plambeck}, {Rosolowsky}, \&
  {Blitz}}]{2003ApJS..149..343E}
{Engargiola}, G., {Plambeck}, R.~L., {Rosolowsky}, E., \& {Blitz}, L. 2003,
  \apjs, 149, 343

\bibitem[{{Engelbracht} {et~al.}(2007){Engelbracht}, {Blaylock}, {Su}, {Rho},
  {Rieke}, {Muzerolle}, {Padgett}, {Hines}, {Gordon}, {Fadda},
  {Noriega-Crespo}, {Kelly}, {Latter}, {Hinz}, {Misselt}, {Morrison},
  {Stansberry}, {Shupe}, {Stolovy}, {Wheaton}, {Young}, {Neugebauer},
  {Wachter}, {P{\'e}rez-Gonz{\'a}lez}, {Frayer}, \&
  {Marleau}}]{2007PASP..119..994E}
{Engelbracht}, C.~W., {Blaylock}, M., {Su}, K.~Y.~L., {et~al.} 2007, \pasp,
  119, 994

\bibitem[{{Engelbracht} {et~al.}(2005){Engelbracht}, {Gordon}, {Rieke},
  {Werner}, {Dale}, \& {Latter}}]{2005ApJ...628L..29E}
{Engelbracht}, C.~W., {Gordon}, K.~D., {Rieke}, G.~H., {et~al.} 2005, \apjl,
  628, L29

\bibitem[{{Fazio} {et~al.}(2004){Fazio}, {Hora}, {Allen}, {Ashby}, {Barmby},
  {Deutsch}, {Huang}, {Kleiner}, {Marengo}, {Megeath}, {Melnick}, {Pahre},
  {Patten}, {Polizotti}, {Smith}, {Taylor}, {Wang}, {Willner}, {Hoffmann},
  {Pipher}, {Forrest}, {McMurty}, {McCreight}, {McKelvey}, {McMurray}, {Koch},
  {Moseley}, {Arendt}, {Mentzell}, {Marx}, {Losch}, {Mayman}, {Eichhorn},
  {Krebs}, {Jhabvala}, {Gezari}, {Fixsen}, {Flores}, {Shakoorzadeh}, {Jungo},
  {Hakun}, {Workman}, {Karpati}, {Kichak}, {Whitley}, {Mann}, {Tollestrup},
  {Eisenhardt}, {Stern}, {Gorjian}, {Bhattacharya}, {Carey}, {Nelson},
  {Glaccum}, {Lacy}, {Lowrance}, {Laine}, {Reach}, {Stauffer}, {Surace},
  {Wilson}, {Wright}, {Hoffman}, {Domingo}, \& {Cohen}}]{2004ApJS..154...10F}
{Fazio}, G.~G., {Hora}, J.~L., {Allen}, L.~E., {et~al.} 2004, \apjs, 154, 10

\bibitem[{{Freedman} {et~al.}(1991){Freedman}, {Wilson}, \&
  {Madore}}]{1991ApJ...372..455F}
{Freedman}, W.~L., {Wilson}, C.~D., \& {Madore}, B.~F. 1991, \apj, 372, 455

\bibitem[{{Gil de Paz} {et~al.}(2007){Gil de Paz}, {Boissier}, {Madore},
  {Seibert}, {Joe}, {Boselli}, {Wyder}, {Thilker}, {Bianchi}, {Rey}, {Rich},
  {Barlow}, {Conrow}, {Forster}, {Friedman}, {Martin}, {Morrissey}, {Neff},
  {Schiminovich}, {Small}, {Donas}, {Heckman}, {Lee}, {Milliard}, {Szalay}, \&
  {Yi}}]{2007ApJS..173..185G}
{Gil de Paz}, A., {Boissier}, S., {Madore}, B.~F., {et~al.} 2007, \apjs, 173,
  185

\bibitem[{{Girardi} {et~al.}(2000){Girardi}, {Bressan}, {Bertelli}, \&
  {Chiosi}}]{2000A&AS..141..371G}
{Girardi}, L., {Bressan}, A., {Bertelli}, G., \& {Chiosi}, C. 2000, \aaps, 141,
  371

\bibitem[{{Greenawalt}(1998)}]{1998PhDT........16G}
{Greenawalt}, B.~E. 1998, PhD thesis, AA(NEW MEXICO STATE UNIVERSITY)

\bibitem[{{Groenewegen} {et~al.}(2007){Groenewegen}, {Wood}, {Sloan},
  {Blommaert}, {Cioni}, {Feast}, {Hony}, {Matsuura}, {Menzies}, {Olivier},
  {Vanhollebeke}, {van Loon}, {Whitelock}, {Zijlstra}, {Habing}, \&
  {Lagadec}}]{2007MNRAS.376..313G}
{Groenewegen}, M.~A.~T., {Wood}, P.~R., {Sloan}, G.~C., {et~al.} 2007, \mnras,
  376, 313

\bibitem[{{Haas} {et~al.}(2002){Haas}, {Klaas}, \&
  {Bianchi}}]{2002A&A...385L..23H}
{Haas}, M., {Klaas}, U., \& {Bianchi}, S. 2002, \aap, 385, L23

\bibitem[{{Herwig}(2005)}]{2005ARA&A..43..435H}
{Herwig}, F. 2005, \araa, 43, 435

\bibitem[{{Heyer} {et~al.}(2004){Heyer}, {Corbelli}, {Schneider}, \&
  {Young}}]{2004ApJ...602..723H}
{Heyer}, M.~H., {Corbelli}, E., {Schneider}, S.~E., \& {Young}, J.~S. 2004,
  \apj, 602, 723

\bibitem[{{Hippelein} {et~al.}(2003){Hippelein}, {Haas}, {Tuffs}, {Lemke},
  {Stickel}, {Klaas}, \& {V{\"o}lk}}]{2003A&A...407..137H}
{Hippelein}, H., {Haas}, M., {Tuffs}, R.~J., {et~al.} 2003, \aap, 407, 137

\bibitem[{{Hirashita} {et~al.}(2007){Hirashita}, {Hibi}, \&
  {Shibai}}]{2007MNRAS.379..974H}
{Hirashita}, H., {Hibi}, Y., \& {Shibai}, H. 2007, \mnras, 379, 974

\bibitem[{{Hoopes} \& {Walterbos}(2000)}]{2000ApJ...541..597H}
{Hoopes}, C.~G. \& {Walterbos}, R.~A.~M. 2000, \apj, 541, 597

\bibitem[{{Issa} {et~al.}(1990){Issa}, {MacLaren}, \&
  {Wolfendale}}]{1990A&A...236..237I}
{Issa}, M.~R., {MacLaren}, I., \& {Wolfendale}, A.~W. 1990, \aap, 236, 237

\bibitem[{{Jarrett} {et~al.}(2003){Jarrett}, {Chester}, {Cutri}, {Schneider},
  \& {Huchra}}]{2003AJ....125..525J}
{Jarrett}, T.~H., {Chester}, T., {Cutri}, R., {Schneider}, S.~E., \& {Huchra},
  J.~P. 2003, \aj, 125, 525

\bibitem[{{Kennicutt}(1989)}]{1989ApJ...344..685K}
{Kennicutt}, Jr., R.~C. 1989, \apj, 344, 685

\bibitem[{{Kennicutt}(1998)}]{1998ARA&A..36..189K}
{Kennicutt}, Jr., R.~C. 1998, \araa, 36, 189

\bibitem[{{Kennicutt} {et~al.}(2007){Kennicutt}, {Calzetti}, {Walter}, {Helou},
  {Hollenbach}, {Armus}, {Bendo}, {Dale}, {Draine}, {Engelbracht}, {Gordon},
  {Prescott}, {Regan}, {Thornley}, {Bot}, {Brinks}, {de Blok}, {de Mello},
  {Meyer}, {Moustakas}, {Murphy}, {Sheth}, \& {Smith}}]{2007ApJ...671..333K}
{Kennicutt}, Jr., R.~C., {Calzetti}, D., {Walter}, F., {et~al.} 2007, \apj,
  671, 333

\bibitem[{{Kroupa}(2001)}]{2001MNRAS.322..231K}
{Kroupa}, P. 2001, \mnras, 322, 231

\bibitem[{{Kruegel}(2003)}]{2003pid..book.....K}
{Kruegel}, E. 2003, {The physics of interstellar dust} (The physics of
  interstellar dust, by Endrik Kruegel.~IoP Series in astronomy and
  astrophysics, ISBN 0750308613.~Bristol, UK: The Institute of Physics, 2003.)

\bibitem[{{Laor} \& {Draine}(1993)}]{1993ApJ...402..441L}
{Laor}, A. \& {Draine}, B.~T. 1993, \apj, 402, 441

\bibitem[{{Leitherer} {et~al.}(1999){Leitherer}, {Schaerer}, {Goldader},
  {Delgado}, {Robert}, {Kune}, {de Mello}, {Devost}, \&
  {Heckman}}]{1999ApJS..123....3L}
{Leitherer}, C., {Schaerer}, D., {Goldader}, J.~D., {et~al.} 1999, \apjs, 123,
  3

\bibitem[{{Magrini} {et~al.}(2007{\natexlab{a}}){Magrini}, {Corbelli}, \&
  {Galli}}]{2007A&A...470..843M}
{Magrini}, L., {Corbelli}, E., \& {Galli}, D. 2007{\natexlab{a}}, \aap, 470,
  843

\bibitem[{{Magrini} {et~al.}(2007{\natexlab{b}}){Magrini}, {V{\'{\i}}lchez},
  {Mampaso}, {Corradi}, \& {Leisy}}]{2007A&A...470..865M}
{Magrini}, L., {V{\'{\i}}lchez}, J.~M., {Mampaso}, A., {Corradi}, R.~L.~M., \&
  {Leisy}, P. 2007{\natexlab{b}}, \aap, 470, 865

\bibitem[{{Makovoz} \& {Marleau}(2005)}]{2005PASP..117.1113M}
{Makovoz}, D. \& {Marleau}, F.~R. 2005, \pasp, 117, 1113

\bibitem[{{Martin} {et~al.}(2005){Martin}, {Fanson}, {Schiminovich},
  {Morrissey}, {Friedman}, {Barlow}, {Conrow}, {Grange}, {Jelinsky},
  {Milliard}, {Siegmund}, {Bianchi}, {Byun}, {Donas}, {Forster}, {Heckman},
  {Lee}, {Madore}, {Malina}, {Neff}, {Rich}, {Small}, {Surber}, {Szalay},
  {Welsh}, \& {Wyder}}]{2005ApJ...619L...1M}
{Martin}, D.~C., {Fanson}, J., {Schiminovich}, D., {et~al.} 2005, \apjl, 619,
  L1

\bibitem[{{Massey} {et~al.}(1995){Massey}, {Armandroff}, {Pyke}, {Patel}, \&
  {Wilson}}]{1995AJ....110.2715M}
{Massey}, P., {Armandroff}, T.~E., {Pyke}, R., {Patel}, K., \& {Wilson}, C.~D.
  1995, \aj, 110, 2715

\bibitem[{{Mathis} {et~al.}(1977){Mathis}, {Rumpl}, \&
  {Nordsieck}}]{1977ApJ...217..425M}
{Mathis}, J.~S., {Rumpl}, W., \& {Nordsieck}, K.~H. 1977, \apj, 217, 425

\bibitem[{{McConnachie} {et~al.}(2006){McConnachie}, {Chapman}, {Ibata},
  {Ferguson}, {Irwin}, {Lewis}, {Tanvir}, \& {Martin}}]{2006ApJ...647L..25M}
{McConnachie}, A.~W., {Chapman}, S.~C., {Ibata}, R.~A., {et~al.} 2006, \apjl,
  647, L25

\bibitem[{{McQuinn} {et~al.}(2007){McQuinn}, {Woodward}, {Willner}, {Polomski},
  {Gehrz}, {Humphreys}, {van Loon}, {Ashby}, {Eicher}, \&
  {Fazio}}]{2007ApJ...664..850M}
{McQuinn}, K.~B.~W., {Woodward}, C.~E., {Willner}, S.~P., {et~al.} 2007, \apj,
  664, 850

\bibitem[{{Miyake} \& {Nakagawa}(1993)}]{1993Icar..106...20M}
{Miyake}, K. \& {Nakagawa}, Y. 1993, Icarus, 106, 20

\bibitem[{{O'Halloran} {et~al.}(2006){O'Halloran}, {Satyapal}, \&
  {Dudik}}]{2006ApJ...641..795O}
{O'Halloran}, B., {Satyapal}, S., \& {Dudik}, R.~P. 2006, \apj, 641, 795

\bibitem[{{P{\'e}rez-Gonz{\'a}lez} {et~al.}(2006){P{\'e}rez-Gonz{\'a}lez},
  {Kennicutt}, {Gordon}, {Misselt}, {Gil de Paz}, {Engelbracht}, {Rieke},
  {Bendo}, {Bianchi}, {Boissier}, {Calzetti}, {Dale}, {Draine}, {Jarrett},
  {Hollenbach}, \& {Prescott}}]{2006ApJ...648..987P}
{P{\'e}rez-Gonz{\'a}lez}, P.~G., {Kennicutt}, Jr., R.~C., {Gordon}, K.~D.,
  {et~al.} 2006, \apj, 648, 987

\bibitem[{{Reach} {et~al.}(2005){Reach}, {Megeath}, {Cohen}, {Hora}, {Carey},
  {Surace}, {Willner}, {Barmby}, {Wilson}, {Glaccum}, {Lowrance}, {Marengo}, \&
  {Fazio}}]{2005PASP..117..978R}
{Reach}, W.~T., {Megeath}, S.~T., {Cohen}, M., {et~al.} 2005, \pasp, 117, 978

\bibitem[{{Regan} \& {Vogel}(1994)}]{1994ApJ...434..536R}
{Regan}, M.~W. \& {Vogel}, S.~N. 1994, \apj, 434, 536

\bibitem[{{Rice} {et~al.}(1990){Rice}, {Boulanger}, {Viallefond}, {Soifer}, \&
  {Freedman}}]{1990ApJ...358..418R}
{Rice}, W., {Boulanger}, F., {Viallefond}, F., {Soifer}, B.~T., \& {Freedman},
  W.~L. 1990, \apj, 358, 418

\bibitem[{{Rieke} {et~al.}(2004){Rieke}, {Young}, {Engelbracht}, {Kelly},
  {Low}, {Haller}, {Beeman}, {Gordon}, {Stansberry}, {Misselt}, {Cadien},
  {Morrison}, {Rivlis}, {Latter}, {Noriega-Crespo}, {Padgett}, {Stapelfeldt},
  {Hines}, {Egami}, {Muzerolle}, {Alonso-Herrero}, {Blaylock}, {Dole}, {Hinz},
  {Le Floc'h}, {Papovich}, {P{\'e}rez-Gonz{\'a}lez}, {Smith}, {Su}, {Bennett},
  {Frayer}, {Henderson}, {Lu}, {Masci}, {Pesenson}, {Rebull}, {Rho}, {Keene},
  {Stolovy}, {Wachter}, {Wheaton}, {Werner}, \&
  {Richards}}]{2004ApJS..154...25R}
{Rieke}, G.~H., {Young}, E.~T., {Engelbracht}, C.~W., {et~al.} 2004, \apjs,
  154, 25

\bibitem[{{Rosolowsky} \& {Simon}(2008)}]{2008ApJ...675.1213R}
{Rosolowsky}, E. \& {Simon}, J.~D. 2008, \apj, 675, 1213

\bibitem[{{Rowe} {et~al.}(2005){Rowe}, {Richer}, {Brewer}, \&
  {Crabtree}}]{2005AJ....129..729R}
{Rowe}, J.~F., {Richer}, H.~B., {Brewer}, J.~P., \& {Crabtree}, D.~R. 2005,
  \aj, 129, 729

\bibitem[{{Salpeter}(1955)}]{1955ApJ...121..161S}
{Salpeter}, E.~E. 1955, \apj, 121, 161

\bibitem[{{Seibert} {et~al.}(2005){Seibert}, {Martin}, {Heckman}, {Buat},
  {Hoopes}, {Barlow}, {Bianchi}, {Byun}, {Donas}, {Forster}, {Friedman},
  {Jelinsky}, {Lee}, {Madore}, {Malina}, {Milliard}, {Morrissey}, {Neff},
  {Rich}, {Schiminovich}, {Siegmund}, {Small}, {Szalay}, {Welsh}, \&
  {Wyder}}]{2005ApJ...619L..55S}
{Seibert}, M., {Martin}, D.~C., {Heckman}, T.~M., {et~al.} 2005, \apjl, 619,
  L55

\bibitem[{{Tabatabaei} {et~al.}(2007{\natexlab{a}}){Tabatabaei}, {Beck},
  {Kr{\"u}gel}, {Krause}, {Berkhuijsen}, {Gordon}, \&
  {Menten}}]{2007A&A...475..133T}
{Tabatabaei}, F.~S., {Beck}, R., {Kr{\"u}gel}, E., {et~al.} 2007{\natexlab{a}},
  \aap, 475, 133

\bibitem[{{Tabatabaei} {et~al.}(2007{\natexlab{b}}){Tabatabaei}, {Krause}, \&
  {Beck}}]{2007A&A...472..785T}
{Tabatabaei}, F.~S., {Krause}, M., \& {Beck}, R. 2007{\natexlab{b}}, \aap, 472,
  785

\bibitem[{{Thilker} {et~al.}(2007{\natexlab{a}}){Thilker}, {Bianchi}, {Meurer},
  {Gil de Paz}, {Boissier}, {Madore}, {Boselli}, {Ferguson},
  {Mu{\~n}oz-Mateos}, {Madsen}, {Hameed}, {Overzier}, {Forster}, {Friedman},
  {Martin}, {Morrissey}, {Neff}, {Schiminovich}, {Seibert}, {Small}, {Wyder},
  {Donas}, {Heckman}, {Lee}, {Milliard}, {Rich}, {Szalay}, {Welsh}, \&
  {Yi}}]{2007ApJS..173..538T}
{Thilker}, D.~A., {Bianchi}, L., {Meurer}, G., {et~al.} 2007{\natexlab{a}},
  \apjs, 173, 538

\bibitem[{{Thilker} {et~al.}(2007{\natexlab{b}}){Thilker}, {Boissier},
  {Bianchi}, {Calzetti}, {Boselli}, {Dale}, {Seibert}, {Braun}, {Burgarella},
  {Gil de Paz}, {Helou}, {Walter}, {Kennicutt}, {Madore}, {Martin}, {Barlow},
  {Forster}, {Friedman}, {Morrissey}, {Neff}, {Schiminovich}, {Small}, {Wyder},
  {Donas}, {Heckman}, {Lee}, {Milliard}, {Rich}, {Szalay}, {Welsh}, \&
  {Yi}}]{2007ApJS..173..572T}
{Thilker}, D.~A., {Boissier}, S., {Bianchi}, L., {et~al.} 2007{\natexlab{b}},
  \apjs, 173, 572

\bibitem[{{Thilker} {et~al.}(2005){Thilker}, {Hoopes}, {Bianchi}, {Boissier},
  {Rich}, {Seibert}, {Friedman}, {Rey}, {Buat}, {Barlow}, {Byun}, {Donas},
  {Forster}, {Heckman}, {Jelinsky}, {Lee}, {Madore}, {Malina}, {Martin},
  {Milliard}, {Morrissey}, {Neff}, {Schiminovich}, {Siegmund}, {Small},
  {Szalay}, {Welsh}, \& {Wyder}}]{2005ApJ...619L..67T}
{Thilker}, D.~A., {Hoopes}, C.~G., {Bianchi}, L., {et~al.} 2005, \apjl, 619,
  L67

\bibitem[{{Verley} {et~al.}(2007){Verley}, {Hunt}, {Corbelli}, \&
  {Giovanardi}}]{2007A&A...476.1161V}
{Verley}, S., {Hunt}, L.~K., {Corbelli}, E., \& {Giovanardi}, C. 2007, \aap,
  476, 1161

\bibitem[{{Weingartner} \& {Draine}(2001)}]{2001ApJS..134..263W}
{Weingartner}, J.~C. \& {Draine}, B.~T. 2001, \apjs, 134, 263

\bibitem[{{Werner} {et~al.}(2004){Werner}, {Roellig}, {Low}, {Rieke}, {Rieke},
  {Hoffmann}, {Young}, {Houck}, {Brandl}, {Fazio}, {Hora}, {Gehrz}, {Helou},
  {Soifer}, {Stauffer}, {Keene}, {Eisenhardt}, {Gallagher}, {Gautier}, {Irace},
  {Lawrence}, {Simmons}, {Van Cleve}, {Jura}, {Wright}, \&
  {Cruikshank}}]{2004ApJS..154....1W}
{Werner}, M.~W., {Roellig}, T.~L., {Low}, F.~J., {et~al.} 2004, \apjs, 154, 1

\bibitem[{{Whitcomb} {et~al.}(1981){Whitcomb}, {Gatley}, {Hildebrand}, {Keene},
  {Sellgren}, \& {Werner}}]{1981ApJ...246..416W}
{Whitcomb}, S.~E., {Gatley}, I., {Hildebrand}, R.~H., {et~al.} 1981, \apj, 246,
  416

\bibitem[{{Whittet}(1992)}]{1992QB791.W45......}
{Whittet}, D.~C.~B. 1992, {Dust in the galactic environment} (Dust in the
  galactic environment Institute of Physics Publishing, 306 p.)

\bibitem[{{Wu} {et~al.}(2005){Wu}, {Cao}, {Hao}, {Liu}, {Wang}, {Xia}, {Deng},
  \& {Young}}]{2005ApJ...632L..79W}
{Wu}, H., {Cao}, C., {Hao}, C.-N., {et~al.} 2005, \apjl, 632, L79

\bibitem[{{Yun} {et~al.}(2001){Yun}, {Reddy}, \&
  {Condon}}]{2001ApJ...554..803Y}
{Yun}, M.~S., {Reddy}, N.~A., \& {Condon}, J.~J. 2001, \apj, 554, 803

\bibitem[{{Zubko} {et~al.}(1996){Zubko}, {Mennella}, {Colangeli}, \&
  {Bussoletti}}]{1996MNRAS.282.1321Z}
{Zubko}, V.~G., {Mennella}, V., {Colangeli}, L., \& {Bussoletti}, E. 1996,
  \mnras, 282, 1321

\end{thebibliography}

\end{document}